\documentclass[12pt]{article}
\usepackage{epsfig,psfig,latexsym,amsbsy,amssymb,cite}                                                          

\def\Spalla{\vbox{\hbox{\null\kern0.25em$\scriptstyle\circ$}%
\vskip-2.70ex\hbox{$S$}}}

\newcommand{\rc}{\right]}
\newcommand{\lc}{\left[}
\newcommand{\ri}{\right)}
\newcommand{\li}{\left(}
\newcommand{\x}{\vec x}

\begin{document}
\setlength{\unitlength}{0.2cm}

\title{Cut-and-permute algorithm for self-avoiding walks in the presence of surfaces}

\author{
  \\
  {\small Maria Serena Causo}             \\[-0.2cm]
  {\small\it INFM-NEST and Scuola Normale Superiore}  \\[-0.2cm]
  {\small\it 56100 Pisa, Italy }          \\[-0.2cm]
  {\small Internet: {\tt causo@sns.it}}     \\[-0.2cm]
  {\protect\makebox[5in]{\quad}}  
  \\
}
\vspace{0.5cm}

\maketitle
\thispagestyle{empty}   

\def\spose#1{\hbox to 0pt{#1\hss}}
\def\ltapprox{\mathrel{\spose{\lower 3pt\hbox{$\mathchar"218$}}
 \raise 2.0pt\hbox{$\mathchar"13C$}}}
\def\gtapprox{\mathrel{\spose{\lower 3pt\hbox{$\mathchar"218$}}
 \raise 2.0pt\hbox{$\mathchar"13E$}}}

\vspace{0.2cm}

\begin{abstract}
We present a dynamic nonlocal hybrid Monte Carlo algorithm consisting of
pivot and ``cut-and-permute'' moves. The algorithm is suitable for the  study 
of polymers in semiconfined geometries at the ordinary transition, where the
pivot algorithm exhibits quasi-ergodic problems.
The dynamic properties of the proposed algorithm are studied in $d = 3$.
The hybrid dynamics is ergodic and exhibits the same optimal critical 
behavior as the pivot algorithm in the bulk.
\end{abstract}
\clearpage

\newcommand{\be}{\begin{equation}}
\newcommand{\ee}{\end{equation}}
\newcommand{\bea}{\begin{eqnarray}}
\newcommand{\eea}{\end{eqnarray}}
\newcommand{\<}{\langle}
\renewcommand{\>}{\rangle}

\def\spose#1{\hbox to 0pt{#1\hss}}
\def\ltapprox{\mathrel{\spose{\lower 3pt\hbox{$\mathchar"218$}}                                           
 \raise 2.0pt\hbox{$\mathchar"13C$}}}
\def\gtapprox{\mathrel{\spose{\lower 3pt\hbox{$\mathchar"218$}}
 \raise 2.0pt\hbox{$\mathchar"13E$}}}                                                                     

\newcommand{\R}{\hbox{{\rm I}\kern-.2em\hbox{\rm R}}}

\newcommand{\reff}[1]{(\ref{#1})}

\section{Introduction}

In this paper we introduce and discuss the properties of a hybrid
Monte Carlo algorithm which can be used to study the equilibrium
properties of a polymer molecule grafted to a surface.

The system can be experimentally obtained either by chemically 
grafting one polymer end in an irreversible way, or, as for surfactants,
via physical adsorption of an endgroup or one of the two blocks
in a diblock copolymer. In this last case, the process is reversible,
since the attached end can desorb both when the temperature increases and
when the solvent quality changes.

Systems of this kind include polymers grafted at colloidal particles
or surfaces in solution which can help stabilizing 
against flocculation~\cite{napper} or polymers
grafted at flexible lipid membranes exerting on the membrane a bending
force which is proportional to the monomer concentration at the 
membrane~\cite{lip1,lip2,lip3,franc}.

In what follows we will consider a polymer in a good solvent 
which is grafted at a surface $S$ and 
interacts repulsively with $S$. 
We will focus in particular on the case in which $S$ is a flat surface.
The polymer will be modelled by an $N$-step lattice self-avoiding walk (SAW), 
which provides a good description of the critical behavior of polymer molecules
in the bulk as well as in confined geometries~\cite{Eisen_book,Diehl_rev}. 

Efficient simulations of SAWs can be obtained by using 
nonlocal algorithms. For instance, the pivot algorithm (see 
Refs.  \cite{Sokal1,Li,Madras-Slade,Alan_Losanna} and references therein) 
is optimal, up to a constant factor, for sampling global observables in the fixed-$N$
free-endpoint ensemble in the bulk since the autocorrelation time in
CPU units is simply proportional to $N$. In the presence of surfaces,
this algorithm is not as efficient and in some cases it is not even 
ergodic (this is the case of a two-dimensional strip, see Ref. \cite{noi}).
In the presence of a single impenetrable plane, the algorithm is 
ergodic, but still one expects it to be inefficient since the 
initial part of the walk will be rarely updated. 
{For} instance, we will show that, in order to update the 
direction of one of the first links pointing
along the normal vector to the surface, the walk should have 
an extremely unlikely geometrical shape. 
This means that, in order to explore the relevant phase space, 
the algorithm has to go through highly improbable configurations.
Therefore, even if ergodic, the algorithm has an autocorrelation time that
grows rapidly with $N$.

To overcome these problems, we propose here a hybrid algorithm
based on the pivot move and on a cut-and-permute move. This last move has been
introduced in Ref.~\cite{CPS-90} in a hybrid algorithm working in
the fixed-endpoint, variable-$N$ ensemble. It
consists in cutting the walk into two parts and in 
rebuilding it by concatenating the two parts in the opposite order.
As we will discuss, such a move is quite efficient in changing the 
configuration of the walk near the grafted endpoint. Thus, if we combine 
the pivot move and the cut-and-permute move, we obtain an algorithm which does 
not have the quasi-ergodicity problems of the pivot algorithm. 
Also, by a careful implementation, it is possible to obtain the 
same optimal behavior as the pivot algorithm for polymers in the bulk: 
the autocorrelation times of global observables in CPU units 
increase only as $N$. It must be noted that, 
even though we consider here only the case of a flat surface, 
the results should apply to any convex surface, for instance to polymers 
grafted outside a sphere: in this respect, the plane should be the worst case.
Note also that we do not consider here any interaction between the surface 
and the walk other than the excluded-volume interaction.
Nonetheless, the results should also hold in the presence of 
attractive interactions as long as they are sufficiently weak and the 
walk is not absorbed. In this case, it is probably important 
to add local moves (and, perhaps, bilocal moves as defined in Ref. \cite{noi})
to speed up the dynamics near the surface. 
Finally, we want to notice that this algorithm is also needed if we want to
apply the join-and-cut algorithm of Ref. \cite{CPS-92} to walks 
in the presence of a surface. 
Indeed, the ideas that are presented here apply directly to that algorithm,
so that by using the cut-and-permute together with the pivot move, 
one should be able to have a version of the join-and-cut algorithm 
which works reasonably well also in the presence of a surface.

The paper is organized as follows. In Sec. 2 and 3 we consider the 
pivot and the cut-and-permute move and discuss in detail the acceptance 
fraction, paying particular attention to those moves in which 
the pivot or the cutting point is near the grafted endpoint of the 
walk. In Sec. 4 we discuss the implementation of the two moves and 
compute the scaling behavior of the average CPU time for each move.
In Sec. 5 we discuss the full algorithm, showing its 
optimal behavior.

\section{The Pivot move in the presence of a surface}

In this paper we will consider $N$-step self-avoiding walks
on a $d$-dimensional lattice $\mathbb{Z}^d$ in the presence of an excluded 
surface $S$  of equation $S(\x) = 0$, $\x \in \mathbb{Z}^d$.
A SAW $\omega$ is a sequence of lattice points $\{ \omega_0, \dots, \omega_N \}$,
such that $\omega_i$, $\omega_{i+1}$ are lattice neighbours and $\omega_i \ne \omega_j$
for $i \ne j$.
The walk is confined in the outward half-space $S(\x) \ge 0$ with its first 
vertex grafted at $S$ and fixed at position $\omega_0$.
The state space is therefore
\be
\Omega_N^{\rm S}(\omega_0) = \{\omega = \{\omega_0,\dots,\omega_N\} | 
S\left(\omega_i \right) \ge 0 \; , \; 
i=1,\dots,N \;{\rm and}\; \omega_i \ne \omega_j\; , \; i\ne j \}.
\label{space}
\ee

As probability measure on the space $\Omega_N^{\rm S}(\omega_0)$, we will consider
the uniform one, which gives equal weight to every allowed walk,
i.e. $\pi\li \omega \ri = {1\over c^{\rm S}_N}$,
where $c^{\rm S}_N$ is the cardinality of $\Omega_N^{\rm S}(\omega_0)$.
In the following we will assume $S$ to be a $(d-1)$ hyperplane
of equation $z=0$.

In $\Omega_N^S(\omega_0)$ we will first consider the pivot algorithm~\cite{Sokal1,Li}.
The elementary move of the algorithm in the bulk consists of the following steps:

\begin{enumerate}

\item
Given a walk configuration $\omega = \{\omega_0, \dots,\omega_N\}$,
an integer $k \in \{0, \dots , N-1\}$ is chosen at random and
the corresponding monomer $\omega_k$ is taken as {\em pivot point}.

\item
An element $g$ of the symmetry group of the lattice is chosen at random 
with probability $p_g$ and a new walk is built by 
applying it to the part of the chain which follows the {\em pivot point},
using $\omega_k$ as fixed point of the transformation.
A walk $\omega'=\{\omega'_0, \dots, \omega'_N \}$,
where the new monomer coordinates are given by

\be
{\omega'}_i = \cases{\omega_i & for $1\le i \le k$, \cr
    \omega_k + g(\omega_i - \omega_k) & for $k+1\le i \le N \,$ ,}
\ee
is obtained.
\item 
If the walk $\omega'$ is not self-avoiding the move is rejected
and the original walk $\omega$ is kept. Otherwise, $\omega'$ is taken as new 
current walk.
\end{enumerate}

We require $p_g=p_{g^{-1}}$ in order to satisfy detailed
balance. In the bulk the algorithm is ergodic if all
axes reflections and either all $\pi /2$ rotations
or all diagonal reflections are given nonzero probability~\cite{Sokal1}. 
Note that in the bulk, in step 1, we can restrict $k$ to belong to 
$\{1,\ldots,N-1\}$, since pivot moves with pivot $\omega_0$ are 
symmetry transformations. However, this is not the case in the presence 
of a surface, and it is thus important to include $k=0$.

A further requirement has to be satisfied in order to take into account
the excluded region:

\begin{description}
\item{4.} If $\omega' \notin \Omega_N^S(\omega_0)$, i.e. if
$S\left( \omega'_i \right) < 0$ for some $i=1, \dots, N$,
the move is rejected and the old walk is counted again.
\end{description}

Because of the presence of the excluded region, ergodicity is not always
satisfied. If $S$ is  
a $(d-1)$-dimensional hyperplane, one can prove that axes reflections and
$\pi /2$ rotations are enough to ensure ergodicity.
However, even if ergodicity is satisfied, one may be worried by the fact
that in order to explore the relevant phase space, the algorithm has to go
through highly improbable configurations. For instance, if we denote by $\hat z$
the direction perpendicular to the hyperplane $S$, we expect the algorithm to
be quite slow in updating the links at the beginning of the walk that are directed
in the $\hat z$ direction.
Indeed, transformations that have the pivot near the surface and that modify
the $\hat z$ direction---moves involving the inversion of the $z$-axis, 
reflections
with respect to the diagonals, $\pi/2$ rotations in the $x=0$ or 
$y=0$ planes---are unlikely to be accepted because of the excluded region.

Therefore, it is important to study the dynamic behaviour 
of the algorithm as a function of the pivot point
in order to determine whether the algorithm is still efficient.
For a generic move based on a lattice symmetry
transformation $g$ applied in the pivot point $\omega_k$
we can define
a local acceptance fraction $f_{g,piv}(N,k)$.
In the scaling limit $N \to \infty$, $k \to \infty$ with $k/N$ finite,
the local acceptance fraction is expected to satisfy the scaling
Ansatz
\be
f_{g,piv}(N,k) \approx N^{-p_{g,piv}} h_{g, piv} \left(k/N \right),
\label{spostata}
\ee
where $p_{g,piv}$ is the exponent which asymptotically governs the decay of the global 
acceptance fraction for $N \to \infty$ and $h_{g, piv} \left(k/N \right)$
is a scaling function.
For small values of $\alpha \equiv k/N$ the behaviour of the 
scaling function $h_{g, piv}(\alpha)$
depends drastically on whether or not the transformation $g$ preserves the
$z$ coordinate of the walk.

In order of analyse quantitatively the dynamic behaviour of the algorithm,
we group the pivot moves into equivalence classes. 
Two pivot moves which are based on the symmetries $g$ and $g'$ 
are said to be equivalent if 
there exists a global symmetry $\sigma$---i.e. a transformation of 
the whole walk---which preserves the geometry of the 
system (i.e. a lattice symmetry which does not modify the $z$ coordinates) 
such that, for every walk $\omega \in \Omega^S(\omega_0)$ and for every 
pivot point $\omega_k$, 
the walk which is obtained by applying a pivot move based on $g'$ to the
globally $\sigma$-transformed walk $\sigma (\omega)$ in the pivot point
$\omega_k' = \sigma(\omega_k)$ and then transforming back via $\sigma^{-1}$
coincides with
the walk obtained from $\omega$ by means of a pivot move based on 
the symmetry $g$ with pivot point $\omega_k$.
In formulae, if we denote by $G_k$ the operator associated with the pivot move
based on $g$ applied on monomer $k$, then
$g$ and $g'$ are equivalent, if, for every $k$,
$G_k = \sigma^{-1} \circ G'_k \circ \sigma$. 
For instance, consider the transformations
$g: (x,y,z) \to (x,z,y)$ and $g': (x,y,z) \to (x,-z,-y)$.
It is a simple matter to show that  they are equivalent:
It is enough to consider 
the $\pi/2$ rotation $\sigma : (x,y,z) \to (x, -z,y)$.

In $3$ dimensions the original $47$ different moves are 
classified in $15$ different classes. 
Moves belonging to a given class have exactly the same critical behaviour.
Therefore, in the following we will study Eq.~\reff{spostata} for different classes
of moves.

\begin{figure}
\centerline{\psfig{file=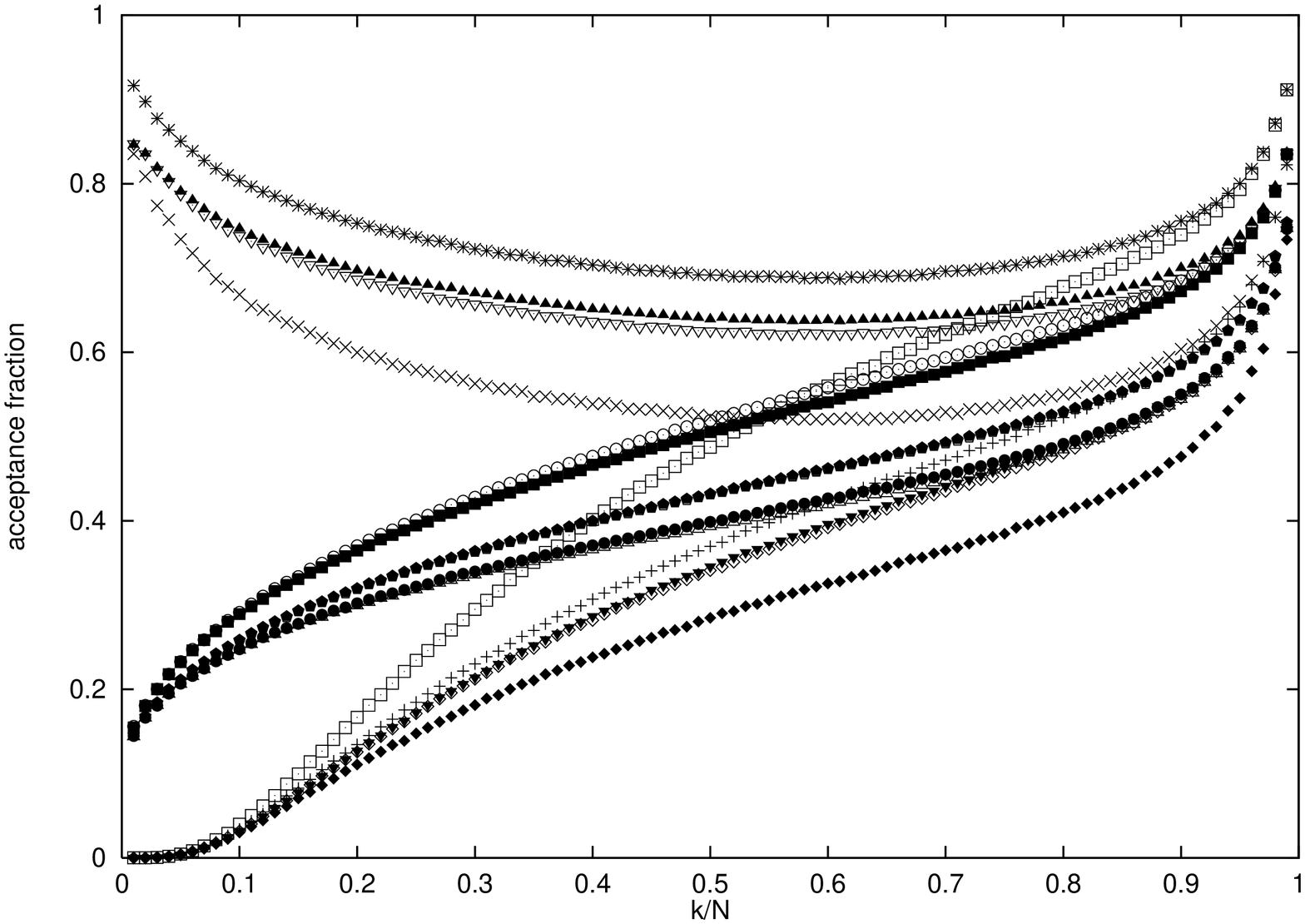,width=0.7\linewidth,angle=-90}}
\caption{Acceptance fraction for the different classes of pivot moves
as a function of the rescaled variable $\alpha = k/N$, for $N =100$.
Three kinds of behaviour are visible at small values of $\alpha$. 
At $\alpha = 0$ the acceptance fraction vanishes for 5 classes of moves transforming 
$z \to -z$, it is small for 6 classes corresponding to $\pi/2$ rotations and
diagonal reflections in the $(y,z)$ and $(z,x)$ planes, while it is not affected
by the presence of the surface for the remaining 4 classes of moves which do
not modify the $z$ coordinate.
The correspondence between symbols and classes of lattice symmetries as defined in 
Sec. 2 is the following:
1.(a) $z$-axis inversion ($\square$), 
1.(b) $x$  or $y$-axis inversion ($\ast$), 
2.(a) $\pm \pi/2$ rotation in $yz$ [or $zx$] planes ($\blacksquare$), 
2.(b) $\pm \pi/2$ rotation in the $xy$ plane ($\triangledown$),
3.(a) $\pi$ rotation in $yz$ [or $zx$] planes ($+$), 
3.(b) $\pi$ rotation in the $xy$ plane ($\times$), 
4.(a) diagonal reflection in the $yz$ [or $zx$] planes ($\circ$), 
4.(b) diagonal reflection in the $xy$ plane ($\blacktriangle$), 
5.(a) diagonal reflection in the $yz$ [or $zx$] plane and $x$ 
[resp. $y$] axis refl. ($\bullet$),
5.(b) diagonal reflection in the $xy$ plane and 
$z$-axis inversion ($\blacktriangledown$), 
6.(a) $\pm \pi/2$ rotation in the $yz$ [or $zx$] planes 
and $x$ [resp. $y$] axis refl.($\vartriangle$), 
6.(b) $\pm \pi/2$ rotation in the $xy$ plane and $z$-axis inversion ($\lozenge$),
7. 3-axes reflection ($\blacklozenge$),
8. diagonal reflection in the $yz$ or $zx$ planes 
and $\pm \pi/2$ rotation in the $xy$ plane (pentagons), 
9. two diagonal reflections, one in the $yz$ or in the $zx$ plane and
the other in the $xy$ plane (full pentagons).
}
\label{fig1}
\end{figure}

In Fig.~\ref{fig1} we report the acceptance fraction for
the 15 classes as a function of $\alpha$, for $N = 100$.
For small $\alpha$, that is for pivot points that are 
near the surface, there are
essentially three types of behaviour.
Moves that involve the inversion of the $z$-axis have a very low acceptance 
rate:
for $\alpha = 0.1$ the acceptance of, say, reflections with respect to the $z$-axis
is $0.04$ and it drops further to $0.0045$ for $\alpha = 0.05$.
Transformations that involve $\pi/2$ rotations and diagonal reflections have a 
better behaviour although the acceptance still drops as $\alpha \to 0$.
Transformations that do not modify the $z$-direction are instead unaffected
by the presence of the surface.

Although the behaviour of the acceptance fraction as a function 
of $\alpha$ is radically different between SAW in the bulk and SAW in a 
half-space, the critical behaviour of the global acceptance fraction averaged
over $k$ is very similar.
  
Denoting with $f_{i,piv}(N,k)$ the acceptance fraction of the class
of moves $i$ applied in the pivot point $\omega_k$,
and with $f_N^{i,piv}$ its average over $k$, we expect $f_N^{i,piv}$ to vanish
as the length of the walk increases as 
\be
f^{i,piv}_N = \frac{1}{N} \sum_{k=0}^{N-1} f_{i,piv}(N,k) \sim N^{-p_{i,piv}} \li A_i + B_i/N^{\Delta_i} + \dots \ri,
\label{subl}
\ee
where the index $i$ indicates a symmetry class, $\Delta_i$ is
the corresponding leading correction-to-scaling exponent, and $A_i$, $B_i$ are
non-universal amplitudes.

Before giving a quantitative estimate
of the exponents $p_{i,piv}$, we try to give heuristically a rough
estimate of the relative order of magnitude of the acceptance
exponents in the bulk and in the presence of the surface.
The argument which we use was already introduced in Ref.~\cite{Sokal1}
for SAW's in the bulk and we simply extend it to the case
of SAW's in a half-space.

If the two parts of the walk $\omega \lc 0,k \rc =
\{ \omega_0, \dots, \omega_k \}$
and $\omega \lc k,N \rc = \{\omega_k, \dots, \omega_N \}$ are
considered as independent, the local acceptance fraction
for any applied lattice symmetry in the bulk would be given by
\be
f_{i,piv}(N,k) \approx \frac{c_N}{c_k c_{N-k}}\, ,
\ee
where $c_k$ is the number of walks of length $k$.
For $k \to \infty$, 
\be c_k \sim \mu^k k^{\gamma -1}\, ,
\label{inbulk}
\ee
where $\mu$ is the critical fugacity and $\gamma \approx 1.1575(5)$ \cite{gamma}, 
so that we have 
\be
f^{i,piv}_N \sim N^{1 - \gamma}\, ,
\ee
and $p_{i,piv} \approx \gamma -1 \approx 0.16$.

In the presence of the surface the same argument gives
\be
f_{i,piv} (N,k) \approx 
\frac{\tilde c_N(0)}{\tilde c_k(0) \tilde c_{N-k}(\omega_k)}\, ,
\label{coipunti}
\ee
where $\tilde c_k(\vec r)$ is the number of walks of length $N$
starting from $\vec r$. For $k \to \infty$, 
$\tilde c_k(0) \sim\mu ^k k^{\tilde \gamma -1}$, 
where $\mu$ is a fugacity which is identical to that appearing in
Eq.~\reff{inbulk} and $\tilde \gamma$ a new critical exponent.
In three dimensions, $\tilde \gamma \approx  0.68$~\cite{grass,Diehl,diehl94}.
Assuming $\omega_k$ to be at a (macroscopically) finite distance
from the surface
we can take $\tilde  c_{N-k} (\omega_k) \propto c_{N-k}$, where $c_{N-k}$
scales according to Eq.~\reff{inbulk}. Then, Eq.~\reff{coipunti} gives again 
$f^{i,piv} \sim N^{1-\gamma}$,
i.e. $p_{i,piv} \approx \gamma - 1 \approx 0.16$.
Thus, heuristically, we expect the acceptance exponents $p_{i,piv}$
to be of the same order in the presence of the surface and in the bulk.
This is confirmed by the numerical estimates.

In order to compute the acceptance exponents $p_i$
we considered walks of length $N = 100$, $200$, $400$, $800$, $4000$,
$8000$, $16000$, $32000$ and performed fits of the global acceptance fraction
for each class of moves to a simple power law.
In order to study the effect of the subleading terms appearing in \reff{subl}
we performed different fits using only data for $ N \ge N_{\rm min}$
for increasing values of $N_{\rm min} = 100, \dots, 16000$. 
We obtained as a result the effective
exponents which are reported in table \ref{tabella_acc_p} and which
approach $p_i$ as the lower cutoff $N_{\rm min}$ increases.

Except for a few cases, which in the following we indicate with a star,
the results are stable for walks of length $N \ge 4000$. 
Our best estimates for the different classes of moves are:
\begin{small}
\begin{enumerate}
\item {(a)} $z$-axis inversion: $p_{piv} = 0.0919(47)\,^\star $,  \\
{(b)}  $x$  or $y$-axis inversion: $p_{piv} = 0.0925(11)$;
\item {(a)} $\pm \pi/2$ rotation in $yz$ [or $zx$] planes: $p_{piv} = 0.10202(81)$,       \\ 
{(b)} $\pm \pi/2$ rotation in the $xy$ plane: $p_{piv} = 0.10002(60)$;    
\item {(a)} $\pi$ rotation in $yz$ [or $zx$] planes: $p_{piv} = 0.1338(12)$,      \\ 
{(b)} $\pi$ rotation in the $xy$ plane: $p_{piv} = 0.1329(17)$;    
\item  {(a)} diagonal reflection in the $yz$ [or $zx$] planes: $p_{piv}= 0.08875(94)\,^\star $,   \\ 
{(b)}  diagonal reflection in the $xy$ plane: $p_{piv} = 0.0887(14)\,^\star $;    
\item {(a)}  diagonal reflection in the $yz$ [or $zx$] plane and $x$ [resp. $y$] axis refl.: $p_{piv} = 0.13024(83)$,\\
{(b)} diagonal reflection in the $xy$ plane and $z$-axis inversion : $p_{piv} = 0.1327(23)$;
\item {(a)}  $\pm \pi/2$ rotation in the $yz$ [or $zx$] planes and $x$ [resp. $y$] axis refl.: $p_{piv} = 0.13556(76)$,\\
{(b)} $\pm \pi/2$ rotation in the $xy$ plane and $z$-axis inversion: $p_{piv} = 0.13371(69)$;     
\item  3-axes reflection: $p_{piv} = 0.1637(17)$;  
\item  diagonal reflection in the $yz$ or $zx$ planes
 and $\pm \pi/2$ rotation in the $xy$ plane: $p_{piv} = 0.11811(77)$;
\item  two diagonal reflections, one in the $yz$ or in the $zx$ plane and 
the other in the $xy$ plane: $p_{piv} = 0.11813(94)$.
\end{enumerate}
\end{small}

Here we have grouped the equivalence classes in the presence
of the surface in 9 groups.
These groups correspond to the 9 equivalence classes for
the algorithm in the bulk, 
where one can consider global transformations $\sigma$
that do not preserve the $z$ coordinate.
{From} the results we observe that the 
exponent $p_{piv}$ depends only on the equivalence
class in the {\em bulk}.
In agreement with the above-reported heuristic argument, 
the presence of the surface 
has no influence on the acceptance exponents of the different classes of moves,
but only on the shape of the scaling functions $h_{i,piv} (\alpha)$
introduced in Eq.~\reff{spostata}.
On the other hand, one should observe that the acceptance exponents which correspond
to different bulk equivalence classes are different, with the only exception
of classes 8. and 9., which can be hardly distinguished even in the
presence of the surface.
One should also notice that the exponent is larger for $3$-axes 
reflections than for diagonal reflections. 
As it has been already noticed in Ref.~\cite{Sokal1} this can
be understood intuitively.
Indeed, two subwalks on the opposite sides
of a pivot point tend to be directed and occupy on average opposite regions of
space. 
We can imagine, for instance, that the two subwalks occupy two opposite octants
which touch at the pivot point. The subwalk $\omega\lc k,N \rc$ is not 
moved in the opposite octant by pivot moves based, for instance, 
on a one-axis reflection,
a diagonal reflection or a $\pi/2$ rotation, while a 3-axes reflection will move
the whole subwalk in the opposite octant. Therefore, it should have a higher probability of
rejection.

The stability of most of the fits with $N_{\rm min}$ seems to rule out
the possibility that all equivalence classes have the same exponent and that 
the discrepancies are due to residual corrections to scaling.
Different symmetries have apparently different acceptance exponents.
If this is correct, the global acceptance exponent $p_{piv}$ averaged over all
transformations would coincide with the
exponent of the pivot move which, in the limit $N \to \infty $, 
has the highest probability
of being accepted, that is $p_{piv} \approx 0.089$.
This exponent is lower than that reported in Ref.~\cite{Li} which was obtained 
by averaging over all transformations. We mention that if we also
perform the group average we obtain $p_{piv} \approx 0.116(1)$.
The group average exponent is only slightly higher than what can be
found using the data reported
in Ref.~\cite{Li} for the pivot algorithm in the bulk,
for which $p \approx 0.113$\footnote{ We have been
informed that the acceptance fractions $f$ reported in Ref.~\cite{Li}
for $d=3$ (but {\em not} for $d=2$) are in error because the program performed 
with probability $1/48$ an identity move (which is always accepted).
Therefore, data there should be corrected by the map
$f \to (48/47)(f - 1/48)$. 
This mistake was guessed by Tom Kennedy on the basis of his own
pivot simulations (see Ref.~\cite{Kennedy})
and was confirmed by Madras and Sokal's examination of their program.
Aware of this problem, we recomputed the acceptance fraction exponent and
found $p \approx 0.113$.}.

Now we can use our best estimates of the acceptance exponents for testing
our scaling Ansatz~\reff{spostata} and determining the behaviour
of the scaling functions $h_{i,piv}(\alpha)$ for $\alpha \equiv k/N \to 0$.
We will show that for $\alpha \to 0$
\be
h_{i,piv}(\alpha) \sim \alpha^{q_i}\, ,
\ee
with $q_i<0$ for the ``good" moves which do not change the $z$ coordinate
of the walk, and $q_i >0$ otherwise.

\begin{figure}
\centerline{\psfig{file=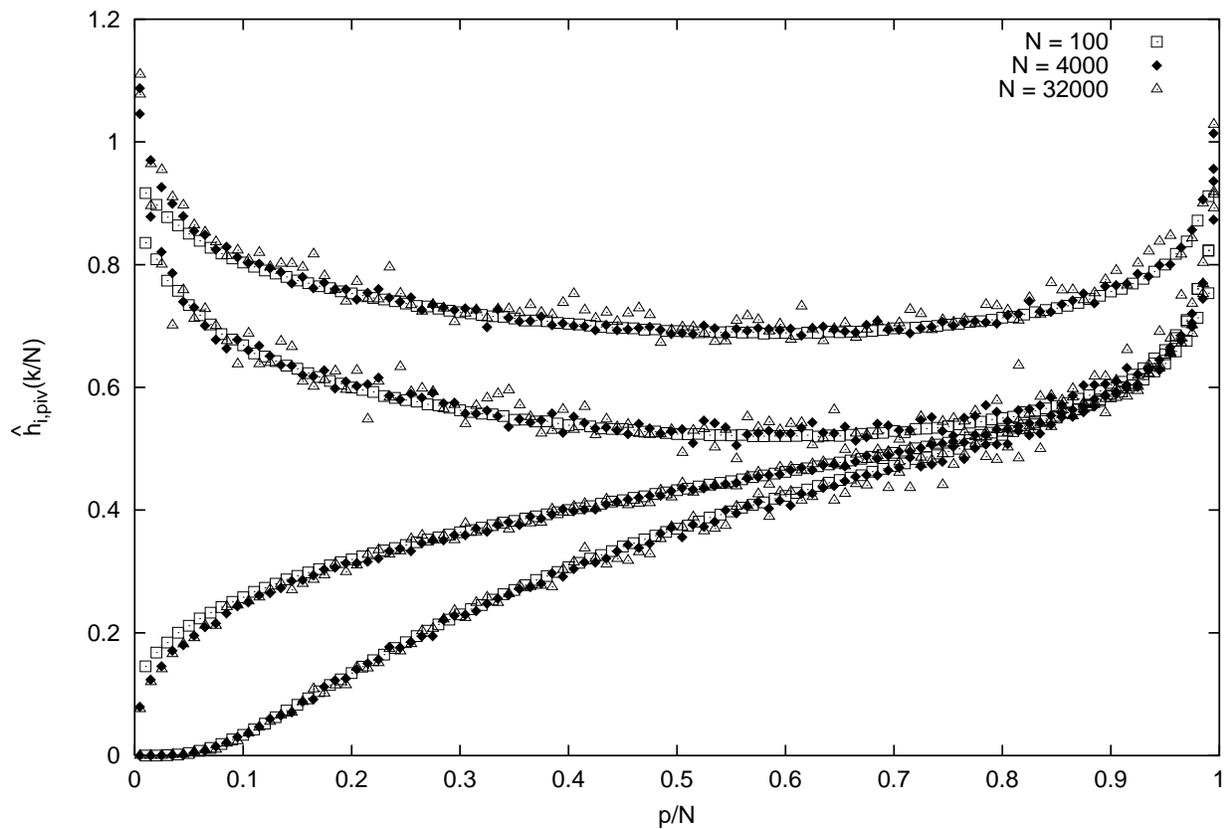,width=0.7\linewidth,angle=-90}}
\caption{Scaling functions in the variable $k/N$ for the pivot acceptance fraction.
The 4 classes of moves considered are, in order of increasing global acceptance:
3.(a), 9., 3.(b) and 1.(b).
}
\label{pivot_risc}
\end{figure}

In Fig.~\ref{pivot_risc} we report $f_{i,piv}(N,k) \cdot (N/100)^{p_{i,piv}} = 
\hat h_{i,piv}(\alpha)$
for 4 different classes of moves: 1.(b), 3.(b), 9. and 3.(a).
They are representative of the different types of behaviour for $\alpha \to 0$.
One observes very good scaling, the data for different $N$ falling one on top
of each other,
except at small values of $\alpha$ for those moves for which $\hat h_{i,piv}$
is nonzero at small $\alpha$ in the range of $N$ values we have considered.
Indeed, for moves 1.(b) and 3.(b) the estimates increase as $N$ increases:
for $N = 100$ we would estimate $\hat h_{{\rm 1.(b)}, piv} (0) \approx 0.92$
and $\hat h_{{\rm 3.(b)}, piv} (0) \approx 0.84$, while for $N = 32000$ 
we have $\hat h_{{\rm 1.(b)}, piv} (0) \approx 1.12$ and 
$\hat h_{{\rm 3.(b)}, piv} (0) \approx 1.10$.
For the moves of class 9., instead, the curve for $N = 100$ is 
slightly higher than that for $N = 4000$, $32000$: for $N = 100$
we would estimate $\hat h_{{\rm 9.},piv}(0) \approx 0.15$, while for $N = 4000$, $32000$
we would obtain $\hat h_{{\rm 9.},piv}(0) \approx 0.07$, although the data for $N=32000$
seem to be even lower. Clearly, for small $\alpha$, there are significant
corrections to scaling and indeed we will now show that 
$ h_{{\rm 1.(b)}, piv} (\alpha)$ and $ h_{{\rm 3.(b)}, piv} (\alpha)$ 
diverge for $\alpha \to 0$, while $ h_{{\rm 9.},piv}(0) =0$.

Consider first moves in classes 1.(b) and 3.(b) and in general all 
the moves preserving the $z$-coordinate of the walks.
They are only marginally affected by the presence of the surface.
The only effect is that, since the surface induces a monomer depletion near the surface,
a move applied in a pivot point $\omega_k$ with small $k$
has higher probability of being accepted than the same move applied in $\omega_{N-k}$,
explaining the slight asymmetry of the scaling curves $h_{i,piv}(\alpha)$.
Moreover, we expect the local acceptance fraction $f_{i,piv}(N,k)$ to remain finite
as $N$ increases with $k$ fixed and small.
By comparing with the scaling Ansatz~\reff{spostata}, it
follows that $h_{i,piv} (\alpha) \sim \alpha^{-p_{i,piv}}$, i.e. 
$q_i = -p_{i,piv}$ and the scaling function
diverges for $\alpha \to 0$. This is confirmed by the numerical results: indeed,
$\hat h_{{\rm 1.(b)},piv} (\alpha)$  and $\hat h_{{\rm 3.(b)},piv} (\alpha)$  
increase as $N \to \infty$.

Let us now consider the classes of moves which change, but do not invert, 
the $z$-coordinate, for instance class 9. 
In order to understand the behaviour of $h_{i,piv} (\alpha)$ for $\alpha \to 0$
we must compute the local acceptance probability for $k$ fixed and small.
For this purpose, we have considered the acceptance fraction 
averaged over the pivot points $k\le k_{\rm max}$, i.e.
\be
\tilde f_{\rm 9. ,piv}(N,k_{\rm max}) = \frac{1}{k_{\rm max}+1} \sum_{k=0}^{k_{\rm max}}
f_{{\rm 9.}, piv}(N,k) \, .
\ee
For fixed $k_{\rm max}$, it decays faster than the global acceptance fraction. For instance, 
for $k_{\rm max} = 20$, for the moves that involve $\pi/2$ rotations
or diagonal reflections in the $(y,z)$, $(z,x)$ planes (classes 8. and 9.), 
we have numerically found 
$\tilde f_{{\rm 8., 9.},,piv}(N,20) \sim N^{-0.49}$. It will be shown in the following
that this exponent is close to the one which characterizes the 
probability that a generic bulk $N$-step SAW grafted at the surface
belongs to $\Omega_N^S (\omega_0)$.
Since we expect that $f_{{\rm 8., 9.},piv}(N,k_{\rm max}) 
\approx \tilde f_{{\rm 8., 9.},piv}(N,k_{\rm max})$
for small $k$, the scaling function $h_{{\rm 8., 9.},piv} (\alpha)$ is expected
to vanish with $q_{\rm 8.,9.}
 \approx 0.49 - p_{{\rm 8., 9.},piv} \approx 0.37$.
Similar behaviour is expected for all classes of moves ( 2.(a), 4.(a),
5.(a), 6.(a), 8. and 9.) which do not invert the $\hat z$ axis.

The behavior of moves like 1.(a) that invert the $z$-axis 
is much worse. For instance, 
we performed a long run for $N=800$, in which such moves were 
attempted on the first 20 monomers approximately 21000 times, and none of the
attempts was successful. 
Fitting directly $\hat h_{{\rm 1.(a)}}(\alpha)$ for $\alpha = k/N \le 0.1$
for $N = 100, 800, 4000$, one finds $q_{\rm 1.(a)} = 3.8(3)$.

It can be useful to get heuristically a rough idea of the order of 
magnitude of the exponent $q_{\rm 1.(a)}$.
Let us denote with $m_N(\omega,k)$ the maximum elongation 
in the $z$ direction of the monomers following the pivot point
\be
m_N(\omega,k) \equiv \max_{i=k+1,\dots,N}{ \omega^z_i},
\ee
where $\omega^z_i$ is the $z$-coordinate of the monomer $\omega_i$.
If we denote with $\omega_k^z$ the $z$-coordinate of the pivot point,
the proposed move does not give rise to collisions with the surface 
if the inequality 
\be
\omega^z_k >\frac{ m_N(\omega,k)}{2}
\label{eur1}
\ee
is satisfied.
Thus, if the pivot point is one of the first vertices following the
grafted end, Eq.~\reff{eur1} states that the move may be successful only for 
those walks that have nearly all their monomers within few lattice planes from the
surface and therefore we expect that $h(\alpha)$ vanishes for $\alpha \to 0$.
It is known (see Ref.~\cite{Eisen_book}) that for 
$N \to \infty$, $z \to \infty$ and 
$\zeta \equiv z/N^\nu$ fixed and small, the probability distribution
$P_N(z)$ of the fraction of monomers which lay at a distance 
$z$ from the surface scales as
\be
P_N(z) \sim \frac{1}{N^\nu} \zeta^{\frac{1}{\nu} - 1}\, .
\label{mon}
\ee
Therefore, the fraction $W(z)$ of monomers that
lay within a distance $z$ from the surface scales as
\be
W(z) = \int_0^z P_N(z') \, dz' \sim \zeta^{1/\nu} \, .
\label{layer}
\ee
If $k$ is small but already in the scaling regime,
we expect that $\omega^z_k \sim k^\nu$ and therefore
$W(2 \omega^z_k) \sim k/N = \alpha$.
Therefore, since the average fraction of monomers laying
in the strip $z < 2 \omega_k^z$ is $\alpha$, the average
fraction of walks such that all their monomers lay in the 
strip is less than $\alpha$. It follows that $q_{\rm 1.(a)} \ge 1$.

Clearly, the pivot move is inefficient when the 
pivot is near the surface. Therefore, we expect the dynamics of observables 
that strongly depend on the behavior of the walk near the surface---for
instance, the number of monomers on the surface---to be much slower than
that of global observables.
Also, note that $\tau_{exp}$ becomes rapidly large,
making it very difficult to thermalize the system using only the pivot
algorithm. In the following Section we discuss a second nonlocal move
that solves the problems we have discussed.

\section{The Cut-and-permute move}

In this Section we want to consider a different nonlocal move that is able to modify
the walk near the surface. By adding it to the pivot algorithm we will 
obtain an algorithm without the quasi-ergodicity problems we have discussed above.
Of course, we do not want to destroy the optimal dynamic behaviour of the pivot
algorithm, and thus we want to introduce a move for which the mean CPU time per succesful
move scales simply with $N$.
The move we introduce here consists in cutting the walk in two parts,
and in rebuilding the walk in reverse order.

The elementary cut-and-permute move works as follows:

\begin{enumerate}
\item[1.] Given an $N$-step SAW $\omega \in \Omega_N^S(\omega_0)$ , choose 
with probability $p_c$ a {\em cut-point} $\omega_c$ with 
$c\in\{1,2,\ldots,N-1\}$. The point $\omega_c$ divides the walk
in two subwalks $\omega^1 = \{ \omega_{1}, \dots, \omega_c\}$ and
${\omega^2} = \{ \omega_{c}, \dots, \omega_N\}$. 
\item[2.] Cut the walk in the {\em cut-point}
and rebuild it arranging the two subwalks $\omega^1$ and ${\omega^2}$  
in reverse order.
The resulting walk $\omega'$ has vertices at positions

\be
{\omega'}_i = \cases{ \omega_0 + \omega_{c+i} - \omega_c & 
      for $0\le i \le N-c$, \cr
              \omega_N- \omega_c + \omega_{i-N+c}& for $N-c+1\le i \le N$.}
\ee
\item[3.]The proposed move is accepted if the resulting walk $\omega'$ is
self-avoiding and does not intersect the excluded region.
\end{enumerate}
It is easy to see that the move satisfies detailed balance as long as 
$p_c = p_{N-c}$.

\begin{figure}
\centerline{\psfig{file=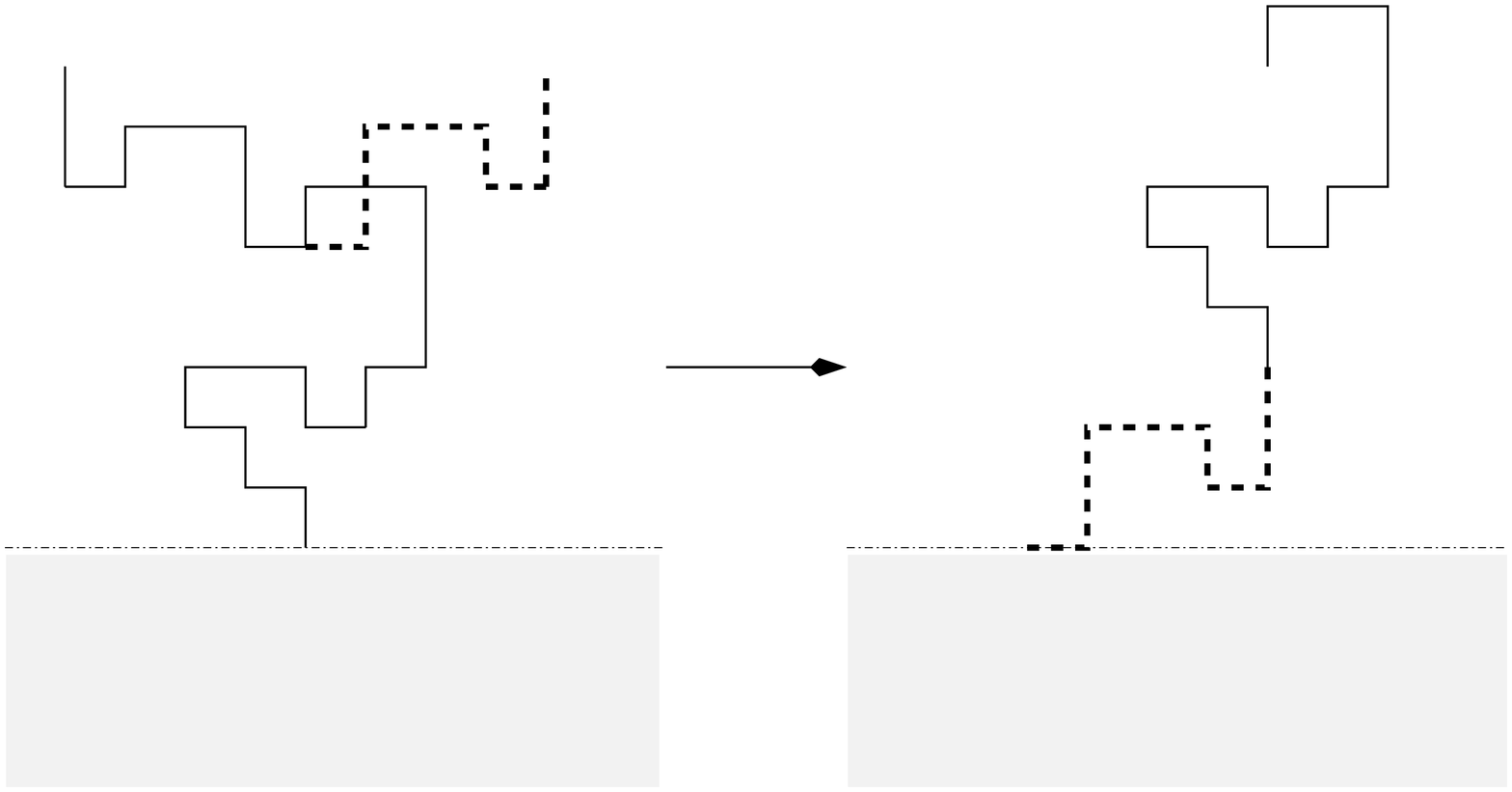,width=0.7\linewidth,angle=-90}}
\caption{A {\em cut-and-permute} move: here $c = 17$ and the symmetry $g$ is an inversion of the 
 $x$-axis.
}
\label{mossa}
\end{figure}

It is also possible to incorporate in the cut-and-permute move the 
``good" pivot transformations, i.e. those that do not change the 
$z$-coordinate. We thus define an ``improved" cut-and-permute move
as follows (the move is illustrated in Fig. \ref{mossa}):
\begin{enumerate}
\item[1.] Given an $N$-step SAW $\omega \in \Omega_N^S$ , choose 
with probability $p_c$ a {\em cut-point} $\omega_c$ with 
$c\in\{1,2,\ldots,N-1\}$. The point $\omega_c$ divides the walk
in two subwalks $\omega^1 = \{ \omega_{1}, \dots, \omega_c\}$ and
${\omega^2} = \{ \omega_{c+1}, \dots, \omega_N\}$. 

\item[2.] Choose with probability $P_g$ an element $g$ belonging to the
symmetry group of the plane which is parallel to the surface.
In three dimensions
the symmetry group is the  dihedral group $D_4$ in
the $z = 0$ plane, whose 8 elements are $\pm \pi/2$ rotations, 
$\pi$ rotation, axis inversions, diagonal reflections, and the identity.

\item[3.] Apply the chosen symmetry to the subwalk $\omega^2$
to obtain the subwalk ${\omega^2}' = \{ \omega_c, \omega_c +
 g(\omega_{c+1} - \omega_c), \dots ,\omega_c + g(\omega_N - \omega_c )\}$.

\item[4.] Cut the walk in the {\em cut-point}
and rebuild it arranging the two subwalks $\omega^1$ and ${\omega^2}'$  
in reverse order.
The resulting walk $\omega'$ has vertices at positions

\be
{\omega'}_i = \cases{ \omega_0 + g(\omega_{c+i} - \omega_c) & for $0\le i \le N-c$, \cr
                      g(\omega_N- \omega_c) + \omega_{i-N+c}& for $N-c+1\le i \le N$.}
\ee
\item[5.]The proposed move is accepted if the resulting walk $\omega'$ is
self-avoiding and does not intersect the excluded region.
\end{enumerate}
The improved cut-and-permute move does not satisfy detailed balance.
Indeed, the walk $\omega'$ obtained from $\omega$ with a cut-and-permute move in 
the cut-point $\omega_c$ cannot be transformed back to the original walk $\omega$ 
by any move applied in $\omega_{N-c}$, since the subwalk of $\omega$ following the
cut-point has changed its orientation in space and its orientation cannot be
modified in the second cut-and-permute move.
However, the move still leaves the probability measure invariant as long
as $p_c = p_{N-c}$. Indeed, if we perform
a move based on $g$ in the point $\omega_{c}$
and then a move based again on $g$ in the point $\omega_{N-c}$,
we obtain the walk $\omega''= g \li \omega \ri$ corresponding to the
application of symmetry $g$
to the original walk. If we denote by $p(\omega \to \omega')$ the 
probability that the move applied to $\omega $ gives as result $\omega'$,
from what we have described it follows that $p(\omega \to \omega') = p(\omega' \to g(\omega))$.
Thus,
\be
\sum_{\omega'} \pi \li \omega' \ri p\li \omega' \to \omega\ri = \sum_{\omega'}\frac{1}{c_N^S} 
p\li g^{-1}( \omega) \to \omega' \ri = \frac{1}{c_N^S}  = \pi \li \omega \ri\, ,
\ee
as required. 
In the following we will consider the improved cut-and-permute move, 
but, as we shall show explicitly, nothing would change by using 
the simpler version. 

As it has been done for the pivot algorithm, we can define equivalence classes of  
cut-and-permute moves and study the acceptance fraction $f_{i,cp}(N,c)$
as a function of the position $c$ of the cut-point
and its average $f^{i,cp}_N$ over all points $c$. For large $N$, we expect 
a scaling behavior of the form
\be 
f^{i,cp}_N \sim N^{-p_{cp}}.
\ee
We have not added an index $i$ to $p_{cp}$ because, as we shall see, 
this exponent does not depend on the equivalence class.

To give an estimate of the exponent $p_{cp}$,
we can use again a heuristic argument in which we consider the two subwalks in which the 
original walk $\omega$ is divided at the cut-point $\omega_c$ as independent.
We also assume that the probability that the attempted move is accepted is the product of the 
probability $P_1(c)$ that the concatenation of the two subwalks gives a walk which 
does not have self-collisions times the probability $P_2(c)$ that the subwalk
$\omega_2$, transformed under a $D_4$ symmetry and translated with its first
vertex on the surface does not intersect the forbidden region.
Repeating the argument presented for the pivot case, we have
\be
P_1(c) = \frac{\tilde c_N}{\tilde c_c  c_{N-c}}\, ,
\ee
which, by averaging over $c$, gives $P_1 \sim N^{-(\gamma-1)} \sim N^{-0.1575(5)} $ .
The second probability is given by 
\be
P_2(c) = \frac{\tilde c_{N-c}}{c_{N-c}} \, ,
\ee
which, after averaging, gives 
$P_2 \sim  N^{-(\gamma -\tilde \gamma)} \sim N^{-0.48}$.
The global acceptance fraction of the cut-and-permute move is expected to scale as
$f_{cp} =P_1 \cdot P_2\sim N^{-p_{cp}} \sim 
N^{-2 \gamma + 1 + \tilde \gamma} \sim N^{-0.63}$.
This argument is likely to give an overestimate of the exponent $p_{cp}$ 
for the following reasons.
The original subwalk $\omega_1$ which is grafted at the surface 
is directed, so that $P_1(c)$ is underestimated.
Moreover, the walk $\omega_2$ is expected to have a residual 
directionality in the $+ \hat z$ direction so that the probability that $\omega_2$
does not intersect the forbidden region is expected to be larger than $P_2(c)$.
Table~\ref{tabella_acc_cp} contains numerical estimates of the acceptances 
$f^{cp,i}_N$ for different 
classes of moves in $D_4$ at different values of $N$,
averaged over the cutting point $c$. 
We have fitted the data
for $N \ge N_{\rm min}$ to a power law
considering increasing values of $N_{\rm min}$. Systematic errors due to corrections 
to scaling should become negligible for $N_{\rm min} \to \infty$.
The $\chi^2$ values of fits with $N_{\rm min} \ge 4000$ indicate that the fits
are stable, with the exception of two cases.
These two cases are denoted with a star in the following list containing our best 
estimates for the exponent $p_{cp}$:
\begin{small}
\begin{enumerate}
\item identity  $p_{cp} = 0.4969(35)$;
\item diagonal reflections  $p_{cp} = 0.4903(29)$;
\item $\pm \pi/2$ rotations $p_{cp} = 0.4911(21)$;
\item $\pi$ rotation  $p_{cp} = 0.4873(40)\,^\star$;
\item 1-axis reflection $p_{cp} = 0.4929(13)\,^\star$.
\end{enumerate}
\end{small}
At variance with the pivot case the results for the different 
equivalence classes are compatible within error bars:
the transformation $g$ seems to play little role. 
If we average over all symmetries we obtain $p_{cp} = 0.4922(20)$.
Note that, as expected, this result is somewhat lower than the heuristic 
estimate reported above.

The exponent $p_{cp}$ is significantly higher than $p_{piv}$
so that cut-and-permute moves will be more rarely accepted.
Numerically, we find that a cut-and-permute move is accepted every 
$\sim 2$ (resp. $\sim 20$) succesful pivot moves for $N=100$ (resp. $N=32000$).
However, they represent the moves that most effectively
change the conformation
of the walk near the grafted end and thus they will play an important role
in decorrelating the walk configurations.

\begin{figure}[ht]
\centerline{\psfig{file=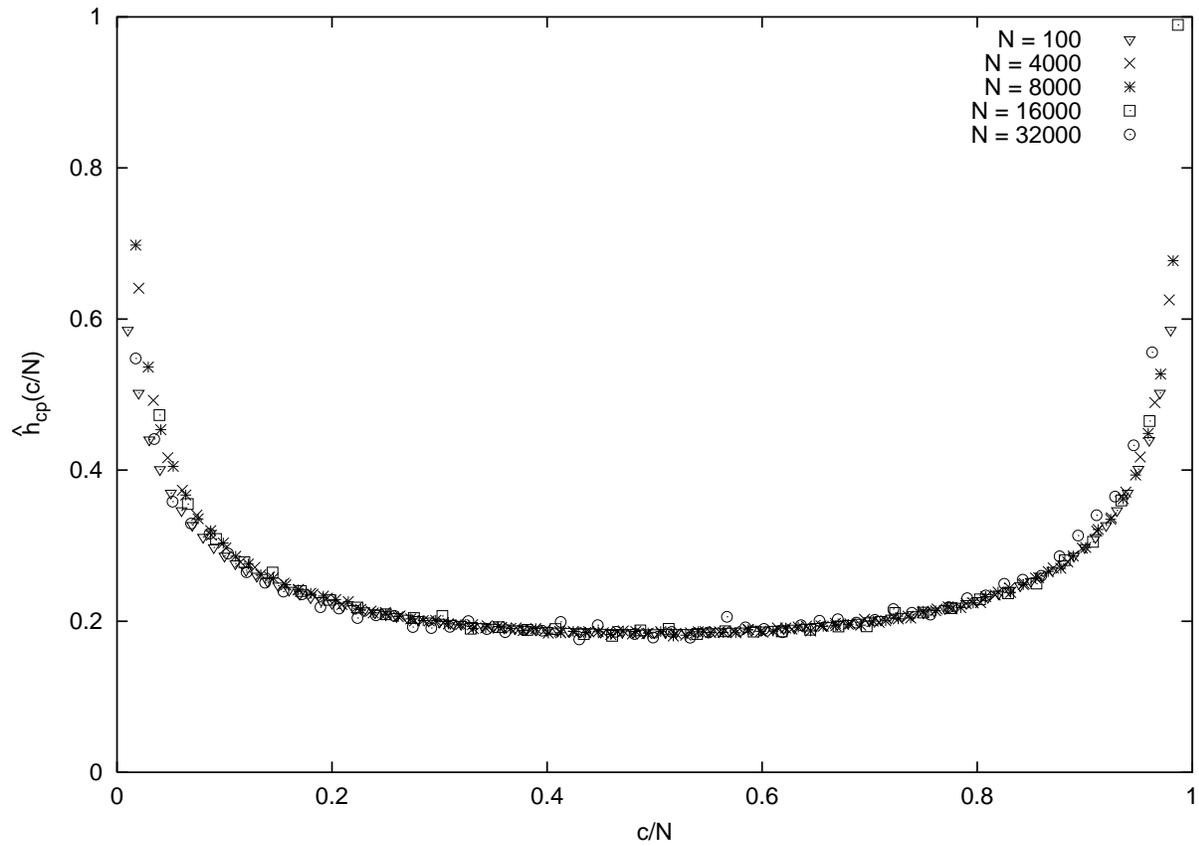,width=0.7\linewidth,angle=-90}}
\caption{
Scaling functions in the variable $c/N$ for the cut-and-permute acceptance fraction.
Data at different $N$ represent the acceptance fraction averaged 
over the different equivalence classes.}
\label{cp_risc}
\end{figure}

As for the pivot case, it is interesting to study the local acceptance 
fraction as a function of $c$ and $N$. For $N \to \infty$, $\zeta\equiv c/N$ 
fixed we expect a scaling form
\be
f_{cp} (N,c) \approx N^{-p_{cp}} h_{cp}(\zeta),
\label{cprisc}
\ee
where $h_{cp}(\zeta)$ is a scaling function which encodes the 
dependence on the different cut-points.
In Figure~\ref{cp_risc} we report the function $\hat h_{cp}(\zeta) =
f_{cp} (N,c) \cdot (N/100)^{p_{cp}}$ for 5
values of $N$ using the exponent $p_{cp}$ given by our fits.
There is no difference in the local behaviour of the 5 classes of equivalent moves,
so the function we report refers to the total acceptance obtained by 
averaging over the equivalence classes. 

Note the very good scaling: the points fall on top of each other for
all values of $N$. 

Let us analyze now the scaling behaviour of the acceptance
at fixed $c$. We define the observable
\be
\tilde f_{cp} (N,c_{\rm max}) = \frac{1}{2 c_{\rm max}} \left[
\sum_{c = 1}^{c_{\rm max}} f_{cp} (N,c)  + 
\sum_{c = N-c_{\rm max}}^{N-1} f_{cp} (N,c) \right]
\, ,
\ee
and study its scaling with $N$ as $N \to \infty$.
For $c_{\rm max} =20$  a power-law fit gives
$\tilde f_{cp} (N,20) \approx N^{-0.01}$. 
This is just an effective exponent and indeed it
decreases in modulus as the minimum value of $N$
considered in the fit is increased. The results are therefore compatible 
with $\tilde f_{cp} (N,20) \approx {\rm const}$
for $N_{\rm min} \ge 4000$. Such a result is easy to understand. 
Indeed, suppose first that $ N-c_{\rm max} \le c < N$. Then the move 
is accepted if $\omega'[0,c']$, $c'=N-c$,
 does not intersect the surface---which happens with 
a probability independent of $N$---and does not intersect 
$\omega'[c',N]$, which, for large $N$, should be roughly independent of 
$N$. Thus, $f_{cp} (N,c)$ should be constant for $N$ large.
If $1\le c \le c_{\rm max}$, note that $f_{cp}(N,c) = f_{cp}(N,N-c)$, 
to conclude the argument. 
In conclusion, the cut-and-permute move has a good scaling behavior exactly in
the region in which the pivot moves behave badly.
The addition of the cut-and-permute move in the hybrid dynamics is therefore
crucial in speeding up the dynamics of the first steps of the
walk.

As in the pivot case, since the local acceptance fraction 
converges to a constant for small or large fixed values of $\zeta$,
the scaling function $h_{cp}(\zeta)$ increases as $\zeta^{-p_{cp}}$ for 
$\zeta \to 0$ (resp. $(1 - \zeta)^{-p_{cp}}$
for $\zeta \to 1$.)
This suggests that in step 1. one can take $p_c$ different from 0 only
for $\zeta$ near 0 and 1, i.e. 
set $p_c = 0$ if $\alpha N \le c \le N (1 - \alpha)$.
For instance, by taking $\alpha = 1/10$, we would approximately increase the
acceptance by a factor $1.7$.

\section{Computational complexity}
\label{cc}

In order to study the effective dynamic behaviour of the algorithm 
in CPU time, it is important to determine the scaling behaviour of the CPU
time needed to generate pivot and cut-and-permute successful moves.

The CPU time depends on the data structure that is used.
For the walk we consider a sequentially allocated linear list and
in order to check for self-collisions and collisions with the
surface we use a hash table as described in Ref.~\cite{Sokal1}.
This allows to insert a single monomer in an average CPU time of order 
one. In the following we will define a CPU time unit as the CPU time
which is needed to upgrade a single monomer position.

In order to optimize the algorithm, it is important to mimimize 
the CPU time spent in a failed attempt.
This requires a careful choice of the order in which the new walk
$\omega'$ is rebuilt.

For the pivot algorithm, Ref.~\cite{Sokal1} suggested to build the 
new walk and perform the self-avoidance check in the following order:
$\omega'_k$, $\omega'_{k+1}$, $\omega'_{k-1}$, $\dots$,$\omega_{k+i}$, $\omega_{k-i}$,
$\dots$ till all monomers have been checked.
The motivation was that the probability of self-intersections is higher 
for those monomers which are close to the pivot point.
However, in the presence of a surface a move that changes the $z$-coordinate
can also fail because the new walk intersects the surface.

Intersections with the surface occur with higher probability for those
monomers of the subwalk $\omega \lc k,N \rc$ which are more likely to
undergo a large displacement, while monomers belonging to the
grafted subwalk $\omega \lc 0,k \rc$ are not modified and satisfy the
geometrical constraint automatically.
In order to keep into account this effect, for the moves that change the 
$z$-coordinates, we can use a different strategy:
we insert the monomers of the proposed walk $\omega'$ in the hash table 
in the following order: $\omega'_k$,
$\omega'_{N}$, $\omega'_{k+1}$, $\omega'_{k-1}$, $\omega'_{N-1}$,
$\dots$, $\omega'_{k+i}$, $\omega'_{k-i}$, $\omega_{N-i}$, $\cdots$ , 
till every monomer has been checked once.
As a measure of the CPU time spent by the algorithm we can use the number of walk
monomers $I(\omega')$ that are checked in the move.
If the move is successful, clearly $I(\omega') = N$, while for a failed attempt
\bea
I(\omega') &=& 
\min \{ i : \omega'\left[k-i,k-1\right]\, \cap\, \omega'\left[k,k+i\right] 
\ne \emptyset \nonumber \\
&&{\rm \, or \,} \omega'\left[N-i,N\right]\,\cap\, \Spalla \ne \emptyset 
{\rm \, or \,} \omega'\left[k-i,k+i\right]\,\cap\, \Spalla \ne \emptyset \}, 
\eea
where, following Ref.~\cite{Sokal1}, we denote the subwalk
$\{\omega'_{\max (i,0)}, \omega'_{\max (i,0)+1}, \cdots \omega'_{\min (j,N)} \}$, 
with the symbol $\omega'\left[i,j\right]$ and 
the half-space $z < 0$ with $\Spalla$.
We want now to evaluate the average time spent in 
checking a failed move that is expected to scale as
\be
T_N^{piv,failed} \equiv \left. \< I(\omega') \> \right|_{\rm failed} \sim
N^{y_{piv}}.
\label{Tpivfailed}
\ee
where the average over all failed moves is taken. An estimate of $y_{piv}$
can be obtained by a heuristic argument.
Inserting monomers in the hash table in the way we described before,
the probability that $I(\omega')$ exceeds the value $i$ can be estimated 
from the scaling of the acceptance fraction. Indeed, we expect
\bea
{\rm Prob} \{I(\omega') > i\}&=&
{\rm Prob} \{\omega'[k-i,k+i] {\rm \, is \, SAW \, and \, 
does \, not \, intersect \, the \, surface},\nonumber\\
 &&{\rm and \, }
\omega'[N-i,N] {\rm \, does \, not \, intersect \, the \, surface \, } \} 
 \sim i^{-p_{piv}}\,.
\eea
Therefore, the average time spent in checking a failed move 
is expected to scale like
\be
T_N^{piv,failed} = \sum_{i=0}^N {\rm Prob} \{I > i \}  \sim N^{1-p_{piv}},
\ee
so that $y_{piv} = 1 - p_{piv}$.
Estimates of $T_N^{piv,failed}$ are reported in Table~\ref{tabella_fail_p},
for the different classes of pivot moves and different values of $N$.
We performed fits of the form \reff{Tpivfailed} in order to 
determine $y_{piv}$, including each time only data with 
$N\ge N_{\rm min}$ (in Table \ref{tabella_fail_p}, $N_{\rm min}$ is the 
value $N$ reported in the first column). 
We observed that in all cases $y_{piv}$ increases with $N_{\rm min}$ 
and thus our results are probably a lower estimate of $y_{piv}$.
Here, we report the results with $N_{\rm min}=4000$ and indicate 
with a star those cases in which the $\chi^2$ is still too large:
\begin{small}
\begin{enumerate}
\item $z$-axis inversion: $y_{piv} = 0.8750(16)$,\\
 $x$  or $y$-axis inversion: $y_{piv}= 0.8234(17)\,^\star $;
\item $\pm \pi/2$ rotation in $yz$ [or $zx$] planes: $y_{piv} = 0.8554(13)^\star $,	\\ 
$\pm \pi/2$ rotation in the $xy$ plane: $y_{piv} = 0.8205(23)$;
\item $\pi$ rotation in $yz$ [or $zx$] planes: $y_{piv} = 0.82756(97)^\star$,	\\ 
$\pi$ rotation in the $xy$ plane: $y_{piv} = 0.7989(46)$;	 
\item  diagonal reflection in the $yz$ [or $zx$] planes: $y_{piv}= 0.8635(22) $,\\	
 diagonal reflection in the $xy$ plane: $y_{piv}= 0.8470(25)^\star$;
\item  diagonal reflection in the $yz$ [or $zx$] plane and $x$ [resp. $y$] axis refl.: $y_{piv}
= 0.8381(13)^\star$,\\
diagonal reflection in the $xy$ plane and $z$-axis inversion : $y_{piv} = 0.8342(16)$;
\item  $\pm \pi/2$ rotation in the $yz$ [or $zx$] planes and $x$ [resp. $y$] axis refl.:
$y_{piv} = 0.8250(18)$,\\
$\pm \pi/2$ rotation in the $xy$ plane and $z$-axis inversion: $y_{piv} = 0.8339(23)$;
\item  3-axis reflection: $y_{piv} = 0.8138(35)$; 
\item  diagonal reflection in the $yz$ or $zx$ planes
 and $\pm \pi/2$ rotation in the $xy$ plane: $y_{piv} = 0.8476(20)^\star$;
\item  two diagonal reflections, one in the $yz$ or in the $zx$ plane and 
the other in the $xy$ plane: $y_{piv} = 0.85121(67)^\star$.
\end{enumerate}
\end{small}
If we average over all equivalence classes we obtain 
$y_{piv} = 0.8428(13)$. These results should not be trusted
too much, and in many cases the correct estimate is probably higher. 
However, the important thing that emerges is that 
$y_{piv} \ltapprox 1 - p_{piv}$, a result that will be used below.

We also computed $T_N^{piv,failed}$ for the moves that change the 
$z$-coordinate when we insert the points as proposed in Ref. \cite{Sokal1},
i.e. without modifying the order in which we insert monomers
into the hash table because of the presence of the surface.
We found that, by using this method, the average number of points 
inserted in the hash
table before detecting a failure is approximately $1.4$ times
larger. The difference is more significant for pivot points $\omega_k$
with $k < N/2$---for $k/N\to 0$ the difference is a factor of two---while
for $k\gtapprox 0.6$ the two methods are equivalent.
However, this has no influence on the exponent $y_{piv}$, which 
is approximately the same in both cases.

At this point we can estimate the average CPU time
spent in a pivot move. Clearly
\bea
T_{piv} &=& (1-f_N^{piv}) T_N^{piv,failed} + f_N^{piv} T_N^{piv,succ}\nonumber \\
&\sim& 1 \cdot N^{y_{piv}} + N^{-p_{piv}} N \sim N^{1-p_{piv}}\, ,
\eea
where we have taken into account that $y_{piv} \ltapprox 1 -p_{piv}$.
Notice also that more time is spent in a successful move than in an
unsuccesful one.

Let us now consider the cut-and-permute move.
As in the previous case we tried to build the new walk in the most efficient 
way in order to reduce the CPU time necessary to detect a failure.
Unlike the pivot case, for the cut-and-permute move the heuristic
argument indicates that the dominant cause of failure is given by 
intersections with the surface.
As it is illustrated in Fig.~\ref{mossa}, the cut-point $\omega_c$
divides the walk in two subwalks, $\omega\lc 0,c \rc$, 
which is grafted at the surface, and $\omega\lc c,N \rc $. 
The move transforms 
$\omega\lc c,N\rc$ via a lattice symmetry $g$ which preserves $z$ 
and translates the transformed subwalk grafting its first vertex at the
surface. The subwalk $\omega[0,c]$ is instead translated in order to join 
the free end of the grafted subwalk.
The only possibility of hitting the surface comes from 
the new grafted subwalk $\omega'\lc 0,N-c\rc$ and the 
monomers which are more likely to intersect
the forbidden region are the ones following the new grafted vertex.
Collision between monomers are instead more likely to occur at the joining
of the two subwalks.

As for the pivot move, in order to check at the same  
time the two regions where failures are more likely to occur,
we inserted the monomers of the proposed new walk $\omega'$ 
in the hash table in the following order: 
$\omega'_{N-c}$, $\omega'_0$, $\omega'_{N-c-1}$,
$\omega'_{N-c+1}$, $\omega'_1$, $\cdots$, 
$\omega'_{i-1}$, $\omega'_{N-c-i}$, $\omega'_{N-c+i}$, $\cdots$.

In this way the CPU time needed to check a proposed walk is
\bea
I(\omega') &=& 
\min \{ i : \omega'\left[N-c-i,N-c-1\right]\, \cap\, \omega'\left[N-c+1,N-c+i\right] 
\ne \emptyset \nonumber \\
&&{\rm \, or \,} \omega'\left[0,i\right]\,\cap\, \Spalla \ne \emptyset 
{\rm \, or \,} \omega'\left[N-c-i,N-c+i\right]\,\cap\, \Spalla \ne \emptyset \}, 
\eea
if $\omega'$ is rejected and $I(\omega') = N$ if it is accepted.
The average time spent in detecting a failed move is expected to scale as 
\be
T_N^{cp,failed} = \left. \< I(\omega')\>\right|_{\rm failed} \sim
N^{y_{cp}}.
\ee
We can estimate $y_{cp}$ heuristically as before.
With the above procedure the probability of using CPU time which is longer than $i$ goes as 
\bea
&&{\rm Prob} \{I(\omega') > i\}=\nonumber\\
&&{\rm Prob} \{\omega'[N-k-i,N-k+i] {\rm \, is \, SAW
\, and \, } \omega'[0,i] {\rm \, does \, not \, intersect \, the \, surface \, } \} \nonumber\\
&& \sim i^{-p_{cp}}\, ,
\eea
and the average total time spent in a failure scales as
\be
T_N^{cp,failed} = \sum_{i=0}^N Prob\{ I>i\} \sim N^{1-p_{cp}}\, ,
\ee
so that $y_{cp} = 1 - p_{cp}$.
The data for $T_N^{cp,failed}$ in the cut-and-permute move and the estimates
of the $y_{cp}$ exponent obtained from data with $N \le N_{min}$, with 
$N_{min}$ appearing in the first column, are 
reported in Table~\ref{tabella_fail_cp}.
{From} those data we have computed our final estimates as we did
in the pivot case obtaining:
\begin{small}
\begin{enumerate}
\item diagonal reflections  $y_{cp} = 0.4870(19)$;
\item identity  $y_{cp} = 0.4788(70)$;
\item $\pm \pi/2$ rotations $y_{cp} = 0.4921(13)$;
\item $\pi$ rotation  $y_{cp} = 0.4933(43)\,^\star$;
\item 1 axis reflection $y_{cp} = 0.4869(13)\,^\star$.
\end{enumerate}
\end{small}
The group average $y_{cp} = 0.4890(15)$ is slightly lower than
$1-p_{cp} \approx 0.5078(20)$ as it is also happened for the pivot
move. 

In order to confirm the role played by intersections
with the surface, we also measured $I(\omega')$ when monomers
are inserted in the hash table in the order
$\omega'_{N-c}$, $\omega'_{N-c+1}$, $\omega'_{N-c-1}$, $\cdots$, without 
inserting at the same time the  monomers near the grafted end.
In this case we find $y_{cp} \approx 0.9$ which is very similar 
to the exponent for the pivot move. The proposed procedure is thus very inefficient.

Finally, we can estimate the average CPU time spent in a 
cut-and-permute move. Clearly 
\bea
T_{N,cp} &=& (1-f_N^{cp})\cdot T_N^{cp,failed} + f_N^{cp}\cdot T_N^{cp,succ} \nonumber \\
&\sim & 1 \cdot N^{y_{cp}} + N^{-p_{cp}}\cdot N \sim N^{1-p_{cp}}\, ,
\eea
where we have taken into account $y_{cp} \ltapprox 1 -p_{cp}$.
Note also that $T_{N,cp}/T_{N,piv} \sim N^{p_{cp} - p_{piv}} \sim N^{-0.4}$,
so that the average CPU time for {\em successful} move scales 
identically.

\section{The hybrid cut-and-permute algorithm: autocorrelation times}

In this Section we want finally to define our hybrid algorithm
that consists in performing both pivot and cut-and-permute moves. 
More precisely, the algorithm is specified by a number $0<q<1$.
The hybrid algorithm works by performing  a pivot move
with probability $q$ and a cut-and-permute move
with probability $(1-q)$.
Since the pivot alone is ergodic, the full algorithm is ergodic.

In order to understand the dynamic behavior we will distinguish three 
different classes of observables: (a) global observables 
that depend on the global shape of the SAW, for instance, the end-to-end
distance or the radius of gyration; (b) global surface observables
that depend on the interaction of the SAW with the surface, 
for example, the number of monomers lying on $S$; (c) local observables
that depend on local properties of the SAW.

Let us analyse at first the behaviour of global observables.
Both pivot and cut-and-permute are nonlocal moves and can
decorrelate global observables in few accepted moves, but the 
cut-and-permute move cannot modify a class of observables,
for instance the $z$-coordinate of the free endpoint.
We denote with $O_z$ the class of observables which  are
updated only in the pivot dynamics and with $O_{x,y}$ all the others.
For a polymer of length $N$ the autocorrelation time $\tau_{int,O_z}$
is therefore of order $1/(q\cdot f_{N,piv})$, while $\tau_{int,O_{x,y}}$ is
of order $1/f_{N,max}$, where $f_{N,max} = \max\{q \cdot f_{N,piv},
(1-q) \cdot f_{N,cp}\}$.
Since in the limit $N \to \infty$ the acceptance fraction of the 
cut-and-permute move decreases faster than the acceptance fraction
of the pivot move, if $q$ is constant, the scaling behaviour of the
autocorrelation time $\tau_{int,glob}$ of any global observable is 
determined by the
pivot dynamics only, i.e. 
\be
\tau_{int,glob} \sim \tau_{int,O_z} \sim \tau_{int,O_{x,y}} \sim
q^{-1} \cdot N^{p_{piv}} \, ,
\label{zeta}
\ee
so that $z_{glob} \approx p_{piv} \approx 0.10$.
Table~\ref{z} reports the autocorrelation times for different values of $N$
for the $z$ component of the free endpoint obtained setting $q = 1/2$.
The dynamic exponent $z $, which is obtained by fitting the data to a power 
law, is $z \approx 0.15$, in reasonable agreement with the prediction.

Global surface observables will be changed in a few successful
cut-and-permute moves and thus we expect 
\be
\tau_{int,surf} \sim (1 - q) N^{p_{cp}},
\ee
so that $z_{surf} \approx p_{cp} \approx 0.49$. Unfortunately, we have not measured
surface quantities in our simulation and therefore we are not able to test
this prediction. Nonetheless, we believe that the correct value of $z$ 
is not too far from $p_{cp}$. 

Finally, let us consider a local observable, for instance the average 
number of monomers at which the SAW makes a 90${}^o$ turn. 
Such a quantity is of order $N$ with a variance of order $\sqrt{N}$. 
By using a standard random-walk argument we obtain
\be
\tau_{int,loc} \sim \min\{q^{-1} N^{1+p_{piv}},(1-q)^{-1} N^{1+p_{cp}}\} 
  \sim N^{1 + p_{piv}} \sim N^{1.09},
\ee
since $p_{cp} < p_{piv}$. Since local moves should be the slowest 
ones we expect $\tau_{exp} \gtapprox N^{1.1}$.

Finally, we estimate the behaviour of the autocorrelation time for 
global observables in CPU units.
Since the average CPU time spent in performing one move of the cut-and-permute
algorithm is proportional to
\be
T = q\cdot T_{piv} + (1-q)\cdot T_{cp} \sim
q\cdot N^{1-p_{piv}} + (1-q)\cdot N^{1-p_{cp}} \sim q \cdot N^{1-p_{piv}}\, ,
\label{tempo}
\ee
the autocorrelation time of global observables in CPU units scales 
as
\bea
\tau^{CPU}_{int,glob} &\sim& N^{z_{glob} + 1 - p_{piv}} \sim N^{1.06}\, , \\
\tau^{CPU}_{int,surf} &\sim& N^{z_{surf} + 1 - p_{piv}} \sim N^{1.40}\, , 
\label{checapp}
\eea
where, in the absence of a numerical estimate, we have used our 
heuristic estimate, $z_{surf}\approx p_{cp}$. The behavior of global
observables is nearly optimal, while for surface observables the 
behavior is not as good. If one is interested in computing surface 
observables, it is possible to improve the algorithm by increasing the 
frequency of the cut-and-permute moves. Indeed, if we scale 
$q^{-1} \sim N^{p_{cp} - p_{piv}}$---essentially we keep the relative 
frequency of successful pivot and cut-and-permute moves constant---then
our previous arguments give 
\be
\tau_{int,glob} \sim N^{p_{cp}}, \qquad\qquad
\tau_{int,surf} \sim N^{p_{cp}}.
\ee
The time per move scales now as $N^{1-p_{cp}}$, so that 
\be
\tau^{CPU}_{int,glob}  \sim
\tau^{CPU}_{int,surf}  \sim N.
\ee
In this case, the dynamics is optimal for both types of observables.

Note that one cannot obtain the same result using the pure pivot
algorithm.
Indeed, the autocorrelation
time for global surface observables would be proportional to the number of
attempts occurring between two accepted pivot moves changing the $z$-coordinate
of the walk and applied in a pivot point $\omega_k$ with $k$ small. 
Because of the monomer depletion near the surface, indeed, one does not
expect that a move at a pivot point far from the grafted end would affect
macroscopically the number of monomers near the surface.
In Sec.2 we have shown that among the $z$-changing moves only the ones
which do not invert $z$ have a small probability of being accepted for small $k$.
This probability decays as $N^{-x}$ with $x \approx p_{cp} \approx 0.49$.
Therefore, if we choose the pivot point with uniform probability, we get
\be
\tau^{\rm pivot}_{int,surf} \sim N^{z^{\rm pivot}_{surf}} \sim N^{1 + p_{cp}},
\label{penultima}
\ee
and the autocorrelation time in CPU units scales as
\be
\tau^{CPU,{\rm pivot}}_{int,surf} \sim N^{z^{\rm pivot}_{surf} + 1 - p_{piv}}\sim N^{2.4}\,.
\label{ultima}
\ee
This indicates clearely that the pivot algorithm alone is inefficient
for simulating polymers grafted at an inpenetrable interface.

\bigskip

{\bf Acknowledgements}

\bigskip

The author is glad to acknowledge stimulating discussions with Sergio Caracciolo,
Peter Grassberger, Andrea Pelissetto, and Stu Whittington.

\clearpage


\clearpage
\begin{table}[ht]
\protect\footnotesize
\begin{center}
\begin{tabular}{|c|c|c|c|c|}
\hline
\hline
$N$ & $f_N^{piv,i} \pm \delta f_N^{piv,i}$& $p_{piv} \pm \delta p_{piv}$ & $\chi^2$ & DF\\
\hline
\hline
1.(a) & \multicolumn{4}{|c|}{ 
 $(x,y,z) \to (x,y,-z)$}\\
\hline

    100. & $      0.44198 \pm       0.00036$ & $     0.09280 \pm       0.00027$ &     15.311 &    6 \\
    200. & $      0.41514 \pm       0.00012$ & $     0.09297 \pm       0.00028$ &     11.529 &    5 \\
    400. & $      0.38911 \pm       0.00070$ & $     0.09266 \pm       0.00054$ &     11.051 &    4 \\
    800. & $      0.36420 \pm       0.00049$ & $     0.09250 \pm       0.00066$ &     10.887 &    3 \\
   4000. & $      0.31525 \pm       0.00051$ & $     0.09572 \pm       0.00151$ &      5.342 &    2 \\
   8000. & $      0.29568 \pm       0.00128$ & $     0.09188 \pm       0.00469$ &      4.592 &    1 \\
  16000. & $      0.27550 \pm       0.00045$ & $     0.07940 \pm       0.00747$ &      0.000 &    0 \\
  32000. & $      0.26075 \pm       0.00128$ & & &\\

\hline
1.(b) & \multicolumn{4}{|c|}{
 $(x,y,z) \to (-x,y,z) \vee (x,-y,z)$}\\
\hline

    100. & $      0.73933 \pm       0.00010$ & $     0.08937 \pm       0.00006$ &    240.326 &    6 \\
    200. & $      0.69716 \pm       0.00018$ & $     0.09035 \pm       0.00009$ &     38.684 &    5 \\
    400. & $      0.65530 \pm       0.00039$ & $     0.09102 \pm       0.00017$ &     17.865 &    4 \\
    800. & $      0.61704 \pm       0.00031$ & $     0.09167 \pm       0.00024$ &      2.573 &    3 \\
   4000. & $      0.53241 \pm       0.00048$ & $     0.09251 \pm       0.00113$ &      1.988 &    2 \\
   8000. & $      0.49970 \pm       0.00012$ & $     0.09511 \pm       0.00216$ &      0.002 &    1 \\
  16000. & $      0.46785 \pm       0.00092$ & $     0.09542 \pm       0.00712$ &      0.000 &    0 \\
  32000. & $      0.43791 \pm       0.00198$ & & &\\

\hline
2.(a) & \multicolumn{4}{|c|}{
$(x,y,z) \to (-z,y,x) \vee (z,y,-x)
\vee (x,-z,y) \vee (x,z,-y)$}\\
\hline

    100. & $      0.49426 \pm       0.00013$ & $     0.10258 \pm       0.00008$ &     92.048 &    6 \\
    200. & $      0.46014 \pm       0.00107$ & $     0.10203 \pm       0.00010$ &      2.764 &    5 \\
    400. & $      0.42768 \pm       0.00005$ & $     0.10202 \pm       0.00010$ &      1.642 &    4 \\
    800. & $      0.39837 \pm       0.00014$ & $     0.10191 \pm       0.00019$ &      1.142 &    3 \\
   4000. & $      0.33805 \pm       0.00021$ & $     0.10202 \pm       0.00081$ &      1.121 &    2 \\
   8000. & $      0.31508 \pm       0.00010$ & $     0.10337 \pm       0.00161$ &      0.188 &    1 \\
  16000. & $      0.29391 \pm       0.00146$ & $     0.10670 \pm       0.00786$ &      0.000 &    0 \\
  32000. & $      0.27296 \pm       0.00062$ & & &\\

\hline
\end{tabular}
\end{center}
\end{table}

\begin{table}[ht]
\protect\footnotesize
\begin{center}
\begin{tabular}{|c|c|c|c|c|}
\hline
2.(b) &\multicolumn{4}{|c|}{
 $(x,y,z) \to (y,-x,z) \vee (-y,x,z)$}\\
\hline

    100. & $      0.67272 \pm       0.00006$ & $     0.09676 \pm       0.00005$ &   1269.317 &    6 \\
    200. & $      0.62958 \pm       0.00065$ & $     0.10012 \pm       0.00011$ &     28.757 &    5 \\
    400. & $      0.59014 \pm       0.00005$ & $     0.10018 \pm       0.00011$ &      6.794 &    4 \\
    800. & $      0.55070 \pm       0.00006$ & $     0.10043 \pm       0.00015$ &      1.166 &    3 \\
   4000. & $      0.46879 \pm       0.00077$ & $     0.10002 \pm       0.00060$ &      0.663 &    2 \\
   8000. & $      0.43679 \pm       0.00025$ & $     0.09978 \pm       0.00067$ &      0.010 &    1 \\
  16000. & $      0.40768 \pm       0.00070$ & $     0.10003 \pm       0.00270$ &      0.000 &    0 \\
  32000. & $      0.38037 \pm       0.00028$ & & &\\

\hline
3.(a) &\multicolumn{4}{|c|}{
 $(x,y,z) \to (x,-y,-z) \vee (-x,y,-z)$}\\
\hline

    100. & $      0.34586 \pm       0.00017$ & $     0.13240 \pm       0.00015$ &    138.561 &    6 \\
    200. & $      0.31323 \pm       0.00014$ & $     0.13106 \pm       0.00019$ &      7.738 &    5 \\
    400. & $      0.28620 \pm       0.00078$ & $     0.13156 \pm       0.00034$ &      4.618 &    4 \\
    800. & $      0.26141 \pm       0.00013$ & $     0.13158 \pm       0.00035$ &      4.535 &    3 \\
   4000. & $      0.21167 \pm       0.00010$ & $     0.13383 \pm       0.00117$ &      0.457 &    2 \\
   8000. & $      0.19300 \pm       0.00021$ & $     0.13462 \pm       0.00269$ &      0.350 &    1 \\
  16000. & $      0.17556 \pm       0.00049$ & $     0.13054 \pm       0.00740$ &      0.000 &    0 \\
  32000. & $      0.16037 \pm       0.00069$ & & &\\

\hline
3.(b) &\multicolumn{4}{|c|}{
 $(x,y,z) \to (-x,-y,z)$}\\
\hline

    100. & $      0.58549 \pm       0.00055$ & $     0.13128 \pm       0.00016$ &     37.268 &    6 \\
    200. & $      0.53739 \pm       0.00008$ & $     0.13145 \pm       0.00016$ &      6.018 &    5 \\
    400. & $      0.49089 \pm       0.00020$ & $     0.13147 \pm       0.00036$ &      6.014 &    4 \\
    800. & $      0.44774 \pm       0.00010$ & $     0.13086 \pm       0.00046$ &      1.565 &    3 \\
   4000. & $      0.36295 \pm       0.00032$ & $     0.13291 \pm       0.00172$ &      0.044 &    2 \\
   8000. & $      0.33084 \pm       0.00094$ & $     0.13242 \pm       0.00358$ &      0.020 &    1 \\
  16000. & $      0.30205 \pm       0.00158$ & $     0.13367 \pm       0.00964$ &      0.000 &    0 \\
  32000. & $      0.27532 \pm       0.00114$ & & &\\

\hline
\end{tabular}
\end{center}
\end{table}

\begin{table}[ht]
\protect\footnotesize
\begin{center}
\begin{tabular}{|c|c|c|c|c|}
\hline
4.(a) &\multicolumn{4}{|c|}{
 $(x,y,z) \to (x,z,y) \vee (x,-z,-y)
\vee (z,y,x) \vee (-z,y,-x)$}\\
\hline

    100. & $      0.50437 \pm       0.00017$ & $     0.09358 \pm       0.00010$ &    165.230 &    6 \\
    200. & $      0.47158 \pm       0.00072$ & $     0.09278 \pm       0.00012$ &     40.909 &    5 \\
    400. & $      0.44132 \pm       0.00008$ & $     0.09277 \pm       0.00012$ &     39.705 &    4 \\
    800. & $      0.41398 \pm       0.00001$ & $     0.09291 \pm       0.00014$ &     34.737 &    3 \\
   4000. & $      0.35583 \pm       0.00027$ & $     0.09094 \pm       0.00049$ &     16.959 &    2 \\
   8000. & $      0.33377 \pm       0.00039$ & $     0.08875 \pm       0.00094$ &      9.538 &    1 \\
  16000. & $      0.31270 \pm       0.00022$ & $     0.08538 \pm       0.00144$ &      0.000 &    0 \\
  32000. & $      0.29473 \pm       0.00021$ & & &\\

\hline
4.(b) &\multicolumn{4}{|c|}{
 $(x,y,z) \to (y,x,z) \vee (-y,-x,z)$}\\
\hline

    100. & $      0.68479 \pm       0.00026$ & $     0.09004 \pm       0.00012$ &    154.206 &    6 \\
    200. & $      0.64473 \pm       0.00098$ & $     0.09132 \pm       0.00016$ &     13.171 &    5 \\
    400. & $      0.60772 \pm       0.00023$ & $     0.09140 \pm       0.00016$ &      6.351 &    4 \\
    800. & $      0.57025 \pm       0.00015$ & $     0.09133 \pm       0.00019$ &      5.922 &    3 \\
   4000. & $      0.49255 \pm       0.00035$ & $     0.09107 \pm       0.00081$ &      5.808 &    2 \\
   8000. & $      0.46179 \pm       0.00022$ & $     0.08866 \pm       0.00141$ &      1.507 &    1 \\
  16000. & $      0.43448 \pm       0.00041$ & $     0.09672 \pm       0.00671$ &      0.000 &    0 \\
  32000. & $      0.40631 \pm       0.00185$ & & &\\

\hline
5.(a) &\multicolumn{4}{|c|}{
$(x,y,z) \to (-x,z,y) \vee (-x,-z,-y)
\vee (z,-y,x) \vee (-z,-y,-x)$}\\
\hline

    100. & $      0.40080 \pm       0.00015$ & $     0.13372 \pm       0.00013$ &    709.276 &    6 \\
    200. & $      0.36448 \pm       0.00073$ & $     0.13174 \pm       0.00015$ &     21.659 &    5 \\
    400. & $      0.32989 \pm       0.00004$ & $     0.13170 \pm       0.00015$ &      4.219 &    4 \\
    800. & $      0.30091 \pm       0.00033$ & $     0.13124 \pm       0.00039$ &      2.576 &    3 \\
   4000. & $      0.24312 \pm       0.00029$ & $     0.13024 \pm       0.00083$ &      0.710 &    2 \\
   8000. & $      0.22245 \pm       0.00042$ & $     0.13062 \pm       0.00179$ &      0.654 &    1 \\
  16000. & $      0.20294 \pm       0.00013$ & $     0.12917 \pm       0.00253$ &      0.000 &    0 \\
  32000. & $      0.18556 \pm       0.00030$ & & &\\

\hline
\end{tabular}
\end{center}
\end{table}

\begin{table}[ht]
\protect\footnotesize
\begin{center}
\begin{tabular}{|c|c|c|c|c|}
\hline
5.(b) &\multicolumn{4}{|c|}{
 $(x,y,z) \to (y,x,-z) \vee (-y,-x,-z)$}\\
\hline

    100. & $      0.32210 \pm       0.00024$ & $     0.13234 \pm       0.00030$ &     46.991 &    6 \\
    200. & $      0.29528 \pm       0.00052$ & $     0.13115 \pm       0.00045$ &     33.813 &    5 \\
    400. & $      0.26675 \pm       0.00019$ & $     0.13021 \pm       0.00048$ &      4.807 &    4 \\
    800. & $      0.24413 \pm       0.00015$ & $     0.13098 \pm       0.00063$ &      1.138 &    3 \\
   4000. & $      0.19781 \pm       0.00022$ & $     0.13268 \pm       0.00228$ &      0.532 &    2 \\
   8000. & $      0.18073 \pm       0.00048$ & $     0.13566 \pm       0.00469$ &      0.006 &    1 \\
  16000. & $      0.16458 \pm       0.00098$ & $     0.13657 \pm       0.01264$ &      0.000 &    0 \\
  32000. & $      0.14971 \pm       0.00096$ & & &\\

\hline
6.(a) &\multicolumn{4}{|c|}{
$(x,y,z) \to (-x,z,-y) \vee (-x,-z,y) 
\vee (-z,-y,x) \vee (z,-y,-x)$}\\
\hline

    100. & $      0.39576 \pm       0.00008$ & $     0.13914 \pm       0.00009$ &    149.867 &    6 \\
    200. & $      0.35778 \pm       0.00084$ & $     0.13683 \pm       0.00021$ &      5.301 &    5 \\
    400. & $      0.32505 \pm       0.00030$ & $     0.13683 \pm       0.00021$ &      5.299 &    4 \\
    800. & $      0.29599 \pm       0.00002$ & $     0.13687 \pm       0.00022$ &      3.607 &    3 \\
   4000. & $      0.23732 \pm       0.00012$ & $     0.13556 \pm       0.00076$ &      0.284 &    2 \\
   8000. & $      0.21601 \pm       0.00022$ & $     0.13525 \pm       0.00149$ &      0.226 &    1 \\
  16000. & $      0.19652 \pm       0.00035$ & $     0.13363 \pm       0.00372$ &      0.000 &    0 \\
  32000. & $      0.17913 \pm       0.00034$ & & &\\

\hline
6.(b) &\multicolumn{4}{|c|}{
 $(x,y,z) \to (-y,x,-z) \vee (y,-x,-z)$}\\
\hline

    100. & $      0.31919 \pm       0.00020$ & $     0.13629 \pm       0.00011$ &    108.973 &    6 \\
    200. & $      0.28836 \pm       0.00025$ & $     0.13629 \pm       0.00012$ &    108.963 &    5 \\
    400. & $      0.26430 \pm       0.00003$ & $     0.13642 \pm       0.00012$ &     35.570 &    4 \\
    800. & $      0.23944 \pm       0.00023$ & $     0.13422 \pm       0.00040$ &      2.428 &    3 \\
   4000. & $      0.19272 \pm       0.00010$ & $     0.13371 \pm       0.00069$ &      1.603 &    2 \\
   8000. & $      0.17576 \pm       0.00007$ & $     0.13506 \pm       0.00127$ &      0.002 &    1 \\
  16000. & $      0.16007 \pm       0.00034$ & $     0.13522 \pm       0.00406$ &      0.000 &    0 \\
  32000. & $      0.14575 \pm       0.00027$ & & &\\

\hline
\end{tabular}
\end{center}
\end{table}

\begin{table}[ht]
\protect\footnotesize
\begin{center}
\begin{tabular}{|c|c|c|c|c|}
\hline
7. &\multicolumn{4}{|c|}{
$(x,y,z) \to (-x,-y,-z)$}\\
\hline

    100. & $      0.27382 \pm       0.00011$ & $     0.16839 \pm       0.00019$ &   1141.917 &    6 \\
    200. & $      0.24357 \pm       0.00012$ & $     0.16438 \pm       0.00025$ &    516.155 &    5 \\
    400. & $      0.21512 \pm       0.00000$ & $     0.16223 \pm       0.00027$ &      1.987 &    4 \\
    800. & $      0.19260 \pm       0.00041$ & $     0.16332 \pm       0.00095$ &      0.554 &    3 \\
   4000. & $      0.14818 \pm       0.00015$ & $     0.16370 \pm       0.00174$ &      0.488 &    2 \\
   8000. & $      0.13222 \pm       0.00014$ & $     0.16129 \pm       0.00404$ &      0.054 &    1 \\
  16000. & $      0.11810 \pm       0.00063$ & $     0.15868 \pm       0.01196$ &      0.000 &    0 \\
  32000. & $      0.10580 \pm       0.00067$ & & &\\

\hline
8. &\multicolumn{4}{|c|}{
$(x,y,z) \to (y,z,-x) \vee (y,-z,x) \vee (-y,z,x) \vee (-y,-z,-x)$ }\\
&\multicolumn{4}{|c|}{
$(x,y,z) \to (z,x,-y) \vee (z,-x,y) \vee (-z,x,y) \vee (-z,-x,-y)$ }\\
\hline

    100. & $      0.42766 \pm       0.00006$ & $     0.12071 \pm       0.00007$ &    198.212 &    6 \\
    200. & $      0.39272 \pm       0.00011$ & $     0.11972 \pm       0.00011$ &     66.816 &    5 \\
    400. & $      0.36094 \pm       0.00019$ & $     0.11894 \pm       0.00015$ &     11.374 &    4 \\
    800. & $      0.33180 \pm       0.00008$ & $     0.11872 \pm       0.00017$ &      1.545 &    3 \\
   4000. & $      0.27392 \pm       0.00016$ & $     0.11811 \pm       0.00077$ &      0.871 &    2 \\
   8000. & $      0.25247 \pm       0.00009$ & $     0.11911 \pm       0.00151$ &      0.282 &    1 \\
  16000. & $      0.23265 \pm       0.00043$ & $     0.12131 \pm       0.00442$ &      0.000 &    0 \\
  32000. & $      0.21389 \pm       0.00052$ & & &\\

\hline
9. &\multicolumn{4}{|c|}{
$(x,y,z) \to (y,z,x) \vee (y,-z,-x) \vee (-y,-z,x) \vee (-y,z,-x)$ }\\
&\multicolumn{4}{|c|}{
$(x,y,z) \to (z,x,y) \vee (z,-x,-y) \vee (-z,-x,y) \vee (-z,x,-y)$ }\\
\hline

    100. & $      0.42794 \pm       0.00003$ & $     0.12050 \pm       0.00008$ &    138.509 &    6 \\
    200. & $      0.39344 \pm       0.00081$ & $     0.11872 \pm       0.00017$ &      8.153 &    5 \\
    400. & $      0.36165 \pm       0.00006$ & $     0.11871 \pm       0.00017$ &      7.110 &    4 \\
    800. & $      0.33281 \pm       0.00010$ & $     0.11821 \pm       0.00026$ &      0.022 &    3 \\
   4000. & $      0.27512 \pm       0.00026$ & $     0.11813 \pm       0.00094$ &      0.013 &    2 \\
   8000. & $      0.25351 \pm       0.00017$ & $     0.11821 \pm       0.00138$ &      0.007 &    1 \\
  16000. & $      0.23351 \pm       0.00065$ & $     0.11781 \pm       0.00485$ &      0.000 &    0 \\
  32000. & $      0.21520 \pm       0.00040$ & & &\\

\hline
\hline
\end{tabular}
\end{center}
\caption{Acceptance fraction and acceptance exponent for different equivalence
classes of pivot moves.}
\label{tabella_acc_p}
\end{table}
\begin{table}[ht]
\protect\footnotesize
\begin{center}
\begin{tabular}{|c|c|c|c|c|}
\hline
\hline
$N$ & $f_N^{cp,i} \pm \delta f_N^{cp,i}$& $p_{cp} \pm \delta p_{cp}$ & $\chi^2$ & DF\\
\hline
\hline
1. &\multicolumn{4}{|c|}{
$(x,y) \to (x,y)$}\\
\hline

    100. & $      0.24511 \pm       0.00008$ & $     0.44881 \pm       0.00021$ &   3544.599 &    6 \\
    200. & $      0.18075 \pm       0.00071$ & $     0.47947 \pm       0.00056$ &     81.376 &    5 \\
    400. & $      0.13252 \pm       0.00001$ & $     0.47979 \pm       0.00057$ &     49.763 &    4 \\
    800. & $      0.09563 \pm       0.00012$ & $     0.48524 \pm       0.00105$ &     12.206 &    3 \\
   4000. & $      0.04397 \pm       0.00007$ & $     0.49686 \pm       0.00352$ &      0.218 &    2 \\
   8000. & $      0.03111 \pm       0.00015$ & $     0.49507 \pm       0.00751$ &      0.146 &    1 \\
  16000. & $      0.02214 \pm       0.00019$ & $     0.50161 \pm       0.01870$ &      0.000 &    0 \\
  32000. & $      0.01564 \pm       0.00015$ & & &\\

\hline
2. &\multicolumn{4}{|c|}{
$(x,y) \to (y,x) \vee (-y,-x)$}\\
\hline

    100. & $      0.24517 \pm       0.00001$ & $     0.46012 \pm       0.00023$ &   2179.892 &    6 \\
    200. & $      0.18130 \pm       0.00027$ & $     0.47995 \pm       0.00051$ &    321.318 &    5 \\
    400. & $      0.13259 \pm       0.00015$ & $     0.48458 \pm       0.00059$ &     65.902 &    4 \\
    800. & $      0.09570 \pm       0.00007$ & $     0.48753 \pm       0.00069$ &      1.363 &    3 \\
   4000. & $      0.04373 \pm       0.00012$ & $     0.49032 \pm       0.00295$ &      0.415 &    2 \\
   8000. & $      0.03116 \pm       0.00006$ & $     0.49210 \pm       0.00467$ &      0.176 &    1 \\
  16000. & $      0.02219 \pm       0.00011$ & $     0.49719 \pm       0.01301$ &      0.000 &    0 \\
  32000. & $      0.01572 \pm       0.00012$ & & &\\

\hline
3. &\multicolumn{4}{|c|}{
$(x,y) \to (-y,x) \vee (y,-x)$}\\
\hline

    100. & $      0.24518 \pm       0.00006$ & $     0.46855 \pm       0.00011$ &  16481.760 &    6 \\
    200. & $      0.18142 \pm       0.00039$ & $     0.48350 \pm       0.00016$ &    753.217 &    5 \\
    400. & $      0.13251 \pm       0.00003$ & $     0.48369 \pm       0.00016$ &    633.232 &    4 \\
    800. & $      0.09576 \pm       0.00004$ & $     0.48822 \pm       0.00024$ &      2.753 &    3 \\
   4000. & $      0.04386 \pm       0.00015$ & $     0.49106 \pm       0.00211$ &      0.915 &    2 \\
   8000. & $      0.03112 \pm       0.00001$ & $     0.49005 \pm       0.00236$ &      0.003 &    1 \\
  16000. & $      0.02217 \pm       0.00021$ & $     0.49086 \pm       0.01458$ &      0.000 &    0 \\
  32000. & $      0.01577 \pm       0.00005$ & & &\\
\hline
\end{tabular}
\end{center}
\end{table}

\begin{table}[ht]
\protect\footnotesize
\begin{center}
\begin{tabular}{|c|c|c|c|c|}
\hline
4. &\multicolumn{4}{|c|}{
$(x,y) \to (-x,-y) $}\\
\hline

    100. & $      0.24507 \pm       0.00016$ & $     0.46150 \pm       0.00035$ &   2088.088 &    6 \\
    200. & $      0.18193 \pm       0.00047$ & $     0.48081 \pm       0.00056$ &    153.010 &    5 \\
    400. & $      0.13278 \pm       0.00007$ & $     0.48182 \pm       0.00058$ &     93.048 &    4 \\
    800. & $      0.09579 \pm       0.00007$ & $     0.48739 \pm       0.00083$ &      3.736 &    3 \\
   4000. & $      0.04365 \pm       0.00015$ & $     0.48733 \pm       0.00402$ &      3.736 &    2 \\
   8000. & $      0.03121 \pm       0.00007$ & $     0.49257 \pm       0.00698$ &      2.894 &    1 \\
  16000. & $      0.02234 \pm       0.00014$ & $     0.52741 \pm       0.02164$ &      0.000 &    0 \\
  32000. & $      0.01550 \pm       0.00021$ & & &\\

\hline
5. &\multicolumn{4}{|c|}{
$(x,y) \to (-x,y) \vee (x,-y)$}\\
\hline
    100. & $      0.24513 \pm       0.00012$ & $     0.46950 \pm       0.00015$ &  10482.119 &    6 \\
    200. & $      0.18120 \pm       0.00055$ & $     0.48043 \pm       0.00019$ &   2075.983 &    5 \\
    400. & $      0.13249 \pm       0.00003$ & $     0.48056 \pm       0.00019$ &   2003.417 &    4 \\
    800. & $      0.09554 \pm       0.00000$ & $     0.48857 \pm       0.00026$ &     21.472 &    3 \\
   4000. & $      0.04384 \pm       0.00009$ & $     0.49292 \pm       0.00136$ &     10.893 &    2 \\
   8000. & $      0.03117 \pm       0.00007$ & $     0.49062 \pm       0.00231$ &      9.382 &    1 \\
  16000. & $      0.02207 \pm       0.00002$ & $     0.48175 \pm       0.00371$ &      0.000 &    0 \\
  32000. & $      0.01580 \pm       0.00004$ & & &\\
\hline
\end{tabular}
\end{center}
\caption{Acceptance fraction and acceptance exponents for equivalent
classes of cut-and-permute moves.}
\label{tabella_acc_cp}
\end{table}
\begin{table}[ht]
\protect\footnotesize
\begin{center}
\begin{tabular}{|c|c|c|c|c|}
\hline
\hline
$N$ & $T^{piv,failed}_N \pm \delta T^{piv,failed}_N$& $y_{piv} \pm \delta y_{piv}$ & $\chi^2$ & DF\\
\hline
\hline
1.(a) &\multicolumn{4}{|c|}{
$(x,y,z) \to (x,y,-z)$}\\
\hline

    100. & $     12.48970 \pm       0.01355$ & $      0.85395 \pm       0.00038$ &   1572.322 &    6 \\
    200. & $     21.88020 \pm       0.06886$ & $      0.86694 \pm       0.00051$ &    116.838 &    5 \\
    400. & $     38.71450 \pm       0.04355$ & $      0.86837 \pm       0.00053$ &     22.278 &    4 \\
    800. & $     70.38660 \pm       0.13222$ & $      0.87022 \pm       0.00079$ &     12.277 &    3 \\
   4000. & $    283.17500 \pm       0.72023$ & $      0.87498 \pm       0.00163$ &      1.072 &    2 \\
   8000. & $    516.57200 \pm       2.50767$ & $      0.87807 \pm       0.00381$ &      0.264 &    1 \\
  16000. & $    954.17300 \pm       8.43733$ & $      0.87159 \pm       0.01318$ &      0.000 &    0 \\
  32000. & $   1745.83000 \pm       3.99927$ & & &\\

\hline
1.(b) &\multicolumn{4}{|c|}{
$(x,y,z) \to (-x,y,z) \vee (x,-y,z)$}\\
\hline

    100. & $     16.83500 \pm       0.00583$ & $      0.78863 \pm       0.00030$ &    818.622 &    6 \\
    200. & $     28.39370 \pm       0.14783$ & $      0.80608 \pm       0.00077$ &    210.977 &    5 \\
    400. & $     47.96640 \pm       0.17382$ & $      0.80994 \pm       0.00087$ &    122.163 &    4 \\
    800. & $     82.41200 \pm       0.31495$ & $      0.81591 \pm       0.00116$ &     61.053 &    3 \\
   4000. & $    299.93200 \pm       0.67181$ & $      0.82335 \pm       0.00166$ &     21.853 &    2 \\
   8000. & $    526.85700 \pm       1.76380$ & $      0.83244 \pm       0.00315$ &     10.320 &    1 \\
  16000. & $    926.25200 \pm       2.84146$ & $      0.84895 \pm       0.00603$ &      0.000 &    0 \\
  32000. & $   1668.36000 \pm       4.73391$ & & &\\

\hline
2.(a) &\multicolumn{4}{|c|}{
$(x,y,z) \to (-z,y,x) \vee (z,y,-x)
\vee (x,-z,y) \vee (x,z,-y)$}\\
\hline

    100. & $     13.90580 \pm       0.00651$ & $      0.82973 \pm       0.00021$ &   2599.558 &    6 \\
    200. & $     24.36420 \pm       0.07489$ & $      0.84812 \pm       0.00042$ &    112.260 &    5 \\
    400. & $     43.11250 \pm       0.11340$ & $      0.84883 \pm       0.00043$ &     45.278 &    4 \\
    800. & $     77.02470 \pm       0.01481$ & $      0.84900 \pm       0.00043$ &     35.698 &    3 \\
   4000. & $    300.49700 \pm       0.49023$ & $      0.85542 \pm       0.00127$ &      6.812 &    2 \\
   8000. & $    540.37900 \pm       1.11283$ & $      0.85976 \pm       0.00212$ &      0.329 &    1 \\
  16000. & $    976.35600 \pm       7.32634$ & $      0.86609 \pm       0.01124$ &      0.000 &    0 \\
  32000. & $   1779.62000 \pm       3.74866$ & & &\\

\hline
\end{tabular}
\end{center}
\end{table}

\begin{table}[ht]
\protect\footnotesize
\begin{center}
\begin{tabular}{|c|c|c|c|c|}
\hline
2.(b) &\multicolumn{4}{|c|}{
 $(x,y,z) \to (y,-x,z) \vee (-y,x,z)$}\\
\hline

    100. & $     14.49390 \pm       0.01393$ & $      0.78652 \pm       0.00038$ &    637.741 &    6 \\
    200. & $     24.40480 \pm       0.06407$ & $      0.79844 \pm       0.00068$ &    193.153 &    5 \\
    400. & $     41.18000 \pm       0.12866$ & $      0.80573 \pm       0.00092$ &     55.520 &    4 \\
    800. & $     71.24760 \pm       0.20412$ & $      0.80964 \pm       0.00125$ &     33.978 &    3 \\
   4000. & $    258.58000 \pm       0.49761$ & $      0.82051 \pm       0.00229$ &      2.083 &    2 \\
   8000. & $    455.17500 \pm       0.91544$ & $      0.82495 \pm       0.00396$ &      0.198 &    1 \\
  16000. & $    807.80900 \pm       3.65702$ & $      0.82059 \pm       0.01058$ &      0.000 &    0 \\
  32000. & $   1426.69000 \pm       8.22899$ & & &\\

\hline
3.(a) &\multicolumn{4}{|c|}{
$(x,y,z) \to (x,-y,-z) \vee (-x,y,-z)$}\\
\hline

    100. & $     11.52150 \pm       0.00774$ & $      0.81006 \pm       0.00012$ &  18940.759 &    6 \\
    200. & $     19.36200 \pm       0.05615$ & $      0.81986 \pm       0.00014$ &   2917.879 &    5 \\
    400. & $     33.04310 \pm       0.03296$ & $      0.82019 \pm       0.00014$ &   2558.777 &    4 \\
    800. & $     57.17680 \pm       0.01268$ & $      0.82104 \pm       0.00015$ &   2056.303 &    3 \\
   4000. & $    212.82400 \pm       0.03815$ & $      0.83872 \pm       0.00044$ &    207.907 &    2 \\
   8000. & $    382.59400 \pm       0.16969$ & $      0.82756 \pm       0.00097$ &     42.738 &    1 \\
  16000. & $    674.83200 \pm       0.71975$ & $      0.84356 \pm       0.00263$ &      0.000 &    0 \\
  32000. & $   1210.97000 \pm       1.79389$ & & &\\

\hline
3.(b) &\multicolumn{4}{|c|}{
$(x,y,z) \to (-x,-y,z)$}\\
\hline

    100. & $     13.49420 \pm       0.01353$ & $      0.74416 \pm       0.00047$ &   4523.560 &    6 \\
    200. & $     21.84060 \pm       0.15401$ & $      0.77124 \pm       0.00062$ &    111.280 &    5 \\
    400. & $     35.90370 \pm       0.00864$ & $      0.77146 \pm       0.00062$ &     82.659 &    4 \\
    800. & $     60.44850 \pm       0.12562$ & $      0.77967 \pm       0.00123$ &     22.250 &    3 \\
   4000. & $    209.65400 \pm       0.73649$ & $      0.79887 \pm       0.00461$ &      3.622 &    2 \\
   8000. & $    363.04100 \pm       0.81002$ & $      0.81292 \pm       0.00905$ &      0.368 &    1 \\
  16000. & $    636.07500 \pm       4.66641$ & $      0.83143 \pm       0.03183$ &      0.000 &    0 \\
  32000. & $   1131.86000 \pm      23.55230$ & & &\\

\hline
\end{tabular}
\end{center}
\end{table}
\begin{table}[ht]
\protect\footnotesize
\begin{center}
\begin{tabular}{|c|c|c|c|c|}
\hline
4.(a) &\multicolumn{4}{|c|}{
$(x,y,z) \to (x,z,y) \vee (x,-z,-y)
\vee (z,y,x) \vee (-z,y,-x)$}\\
\hline

    100. & $     13.55210 \pm       0.00593$ & $      0.83932 \pm       0.00032$ &   1054.187 &    6 \\
    200. & $     23.77720 \pm       0.02946$ & $      0.85113 \pm       0.00051$ &    163.158 &    5 \\
    400. & $     42.22660 \pm       0.05254$ & $      0.85557 \pm       0.00063$ &     26.317 &    4 \\
    800. & $     75.99580 \pm       0.10340$ & $      0.85770 \pm       0.00086$ &     12.732 &    3 \\
   4000. & $    301.14100 \pm       0.61872$ & $      0.86352 \pm       0.00219$ &      4.372 &    2 \\
   8000. & $    544.65900 \pm       1.36295$ & $      0.86954 \pm       0.00363$ &      0.029 &    1 \\
  16000. & $    996.21800 \pm       6.31179$ & $      0.86773 \pm       0.01119$ &      0.000 &    0 \\
  32000. & $   1817.89000 \pm       8.13467$ & & &\\
\hline
4.(b) &\multicolumn{4}{|c|}{
$(x,y,z) \to (y,x,z) \vee (-y,-x,z)$}\\
\hline

    100. & $     13.92810 \pm       0.01710$ & $      0.79980 \pm       0.00034$ &   2770.326 &    6 \\
    200. & $     23.31840 \pm       0.07178$ & $      0.81499 \pm       0.00046$ &    370.328 &    5 \\
    400. & $     40.02870 \pm       0.10436$ & $      0.81686 \pm       0.00048$ &    194.954 &    4 \\
    800. & $     69.26540 \pm       0.04385$ & $      0.81769 \pm       0.00050$ &    145.693 &    3 \\
   4000. & $    256.45000 \pm       0.83364$ & $      0.83938 \pm       0.00209$ &     30.983 &    2 \\
   8000. & $    451.01700 \pm       0.58539$ & $      0.84708 \pm       0.00251$ &      0.871 &    1 \\
  16000. & $    809.00600 \pm       2.64035$ & $      0.85312 \pm       0.00694$ &      0.000 &    0 \\
  32000. & $   1461.39000 \pm       5.17075$ & & &\\

\hline
5.(a) &\multicolumn{4}{|c|}{
$(x,y,z) \to (-x,z,y) \vee (-x,-z,-y)
\vee (z,-y,x) \vee (-z,-y,-x)$}\\
\hline

    100. & $     12.19490 \pm       0.00138$ & $      0.79610 \pm       0.00009$ &  13776.870 &    6 \\
    200. & $     20.54390 \pm       0.04020$ & $      0.81876 \pm       0.00022$ &    977.370 &    5 \\
    400. & $     35.10850 \pm       0.04365$ & $      0.82039 \pm       0.00023$ &    407.213 &    4 \\
    800. & $     60.83820 \pm       0.02967$ & $      0.82179 \pm       0.00025$ &    180.301 &    3 \\
   4000. & $    225.88200 \pm       0.43779$ & $      0.83565 \pm       0.00109$ &     10.462 &    2 \\
   8000. & $    400.64600 \pm       0.27076$ & $      0.83814 \pm       0.00134$ &      0.074 &    1 \\
  16000. & $    716.28200 \pm       0.46864$ & $      0.83497 \pm       0.01175$ &      0.000 &    0 \\
  32000. & $   1277.72000 \pm      10.36998$ & & &\\

\hline
\end{tabular}
\end{center}
\end{table}

\begin{table}[ht]
\protect\footnotesize
\begin{center}
\begin{tabular}{|c|c|c|c|c|}
\hline
5.(b) &\multicolumn{4}{|c|}{
$(x,y,z) \to (y,x,-z) \vee (-y,-x,-z)$}\\
\hline

    100. & $     10.63500 \pm       0.00391$ & $      0.79204 \pm       0.00013$ &   6828.502 &    6 \\
    200. & $     17.72970 \pm       0.02629$ & $      0.81120 \pm       0.00027$ &    621.314 &    5 \\
    400. & $     30.14010 \pm       0.02065$ & $      0.81429 \pm       0.00031$ &    179.047 &    4 \\
    800. & $     52.36940 \pm       0.10904$ & $      0.82540 \pm       0.00101$ &     46.110 &    3 \\
   4000. & $    195.91700 \pm       0.06318$ & $      0.83425 \pm       0.00165$ &      0.147 &    2 \\
   8000. & $    349.47400 \pm       0.61497$ & $      0.83322 \pm       0.00358$ &      0.042 &    1 \\
  16000. & $    621.76000 \pm       4.43418$ & $      0.83566 \pm       0.01243$ &      0.000 &    0 \\
  32000. & $   1109.64000 \pm       5.36154$ & & &\\

\hline
6.(a) &\multicolumn{4}{|c|}{
$(x,y,z) \to (-x,z,-y) \vee (-x,-z,y) 
\vee (-z,-y,x) \vee (z,-y,-x)$}\\
\hline

    100. & $     12.31430 \pm       0.01259$ & $      0.79843 \pm       0.00032$ &   3167.891 &    6 \\
    200. & $     20.63680 \pm       0.05582$ & $      0.80658 \pm       0.00035$ &    228.953 &    5 \\
    400. & $     35.42490 \pm       0.00217$ & $      0.80680 \pm       0.00035$ &    180.458 &    4 \\
    800. & $     61.23020 \pm       0.08125$ & $      0.81473 \pm       0.00075$ &     38.007 &    3 \\
   4000. & $    225.32100 \pm       0.29390$ & $      0.82502 \pm       0.00184$ &      0.293 &    2 \\
   8000. & $    399.46000 \pm       0.66259$ & $      0.82396 \pm       0.00337$ &      0.152 &    1 \\
  16000. & $    705.99200 \pm       3.17945$ & $      0.82742 \pm       0.00948$ &      0.000 &    0 \\
  32000. & $   1252.79000 \pm       5.99259$ & & &\\

\hline
6.(b) &\multicolumn{4}{|c|}{
$(x,y,z) \to (-y,x,-z) \vee (y,-x,-z)$}\\
\hline

    100. & $     10.72070 \pm       0.01194$ & $      0.79288 \pm       0.00041$ &   3124.323 &    6 \\
    200. & $     17.77960 \pm       0.03092$ & $      0.81265 \pm       0.00057$ &    677.940 &    5 \\
    400. & $     30.26860 \pm       0.12081$ & $      0.82221 \pm       0.00069$ &     50.629 &    4 \\
    800. & $     52.47530 \pm       0.04430$ & $      0.82291 \pm       0.00070$ &     25.697 &    3 \\
   4000. & $    195.23300 \pm       0.83783$ & $      0.83388 \pm       0.00233$ &      1.349 &    2 \\
   8000. & $    346.10300 \pm       0.87541$ & $      0.83597 \pm       0.00295$ &      0.001 &    1 \\
  16000. & $    617.91800 \pm       4.27681$ & $      0.83572 \pm       0.01101$ &      0.000 &    0 \\
 32000. & $   1102.83000 \pm       3.55154$ & & &\\
\hline
\end{tabular}
\end{center}
\end{table}

\begin{table}[ht]
\protect\footnotesize
\begin{center}
\begin{tabular}{|c|c|c|c|c|}

\hline
7. &\multicolumn{4}{|c|}{
$(x,y,z) \to (-x,-y,-z)$}\\
\hline

    100. & $     10.43374 \pm       0.00562$ & $      0.72192 \pm       0.00026$ &   6189.621 &    6 \\
    200. & $     16.87000 \pm       0.15782$ & $      0.79349 \pm       0.00097$ &    262.240 &    5 \\
    400. & $     27.59870 \pm       0.08282$ & $      0.79487 \pm       0.00098$ &    164.803 &    4 \\
    800. & $     46.24860 \pm       0.00737$ & $      0.79758 \pm       0.00100$ &     24.244 &    3 \\
   4000. & $    163.67700 \pm       1.03384$ & $      0.81382 \pm       0.00354$ &      1.338 &    2 \\
   8000. & $    285.89400 \pm       1.67814$ & $      0.81769 \pm       0.00538$ &      0.421 &    1 \\
  16000. & $    501.00000 \pm       3.85448$ & $      0.82535 \pm       0.01297$ &      0.000 &    0 \\
  32000. & $    887.75500 \pm       4.12887$ & & &\\

\hline
8. &\multicolumn{4}{|c|}{
$(x,y,z) \to (y,z,-x) \vee (y,-z,x) \vee (-y,z,x) \vee (-y,-z,-x)$ }\\
&\multicolumn{4}{|c|}{
$(x,y,z) \to (z,x,-y) \vee (z,-x,y) \vee (-z,x,y) \vee (-z,-x,-y)$ }\\
\hline

    100. & $     12.46370 \pm       0.00513$ & $      0.80822 \pm       0.00021$ &   7715.820 &    6 \\
    200. & $     21.30210 \pm       0.00692$ & $      0.81994 \pm       0.00026$ &   2238.419 &    5 \\
    400. & $     37.03280 \pm       0.00907$ & $      0.82865 \pm       0.00033$ &    242.017 &    4 \\
    800. & $     65.16220 \pm       0.05450$ & $      0.83399 \pm       0.00052$ &     66.482 &    3 \\
   4000. & $    247.15100 \pm       0.43455$ & $      0.84409 \pm       0.00140$ &      5.681 &    2 \\
   8000. & $    441.53500 \pm       0.69148$ & $      0.84757 \pm       0.00205$ &      0.269 &    1 \\
  16000. & $    792.69500 \pm       3.48873$ & $      0.85116 \pm       0.00723$ &      0.000 &    0 \\
  32000. & $   1429.98000 \pm       3.41729$ & & &\\

\hline
9. &\multicolumn{4}{|c|}{
$(x,y,z) \to (y,z,x) \vee (y,-z,-x) \vee (-y,-z,x) \vee (-y,z,-x)$ }\\
&\multicolumn{4}{|c|}{
$(x,y,z) \to (z,x,y) \vee (z,-x,-y) \vee (-z,-x,y) \vee (-z,x,-y)$ }\\
\hline

    100. & $     12.44090 \pm       0.00677$ & $      0.82676 \pm       0.00012$ &  18185.883 &    6 \\
    200. & $     21.30550 \pm       0.01874$ & $      0.83542 \pm       0.00014$ &   2970.316 &    5 \\
    400. & $     37.06200 \pm       0.06948$ & $      0.83729 \pm       0.00014$ &    715.471 &    4 \\
    800. & $     65.09910 \pm       0.01040$ & $      0.83740 \pm       0.00014$ &    629.169 &    3 \\
   4000. & $    246.75200 \pm       0.26749$ & $      0.84924 \pm       0.00050$ &     24.712 &    2 \\
   8000. & $    442.15800 \pm       0.32010$ & $      0.85121 \pm       0.00067$ &      4.541 &    1 \\
  16000. & $    793.87800 \pm       1.71463$ & $      0.85794 \pm       0.00323$ &      0.000 &    0 \\
  32000. & $   1438.86000 \pm       0.83247$ & & &\\

\hline
\end{tabular}
\end{center}
\caption{Average time spent in a failed pivot move and 
scaling-behaviour exponent $y_{piv}$
for the equivalence classes of pivot moves.
The check of self-avoidance was performed as explained in section~\ref{cc}}
\label{tabella_fail_p}
\end{table}
\begin{table}[ht]
\protect\footnotesize
\begin{center}
\begin{tabular}{|c|c|c|c|c|}
\hline
\hline
N & $T_N^{cp,failed} \pm \delta T_N^{cp,failed}$& $y_{cp} \pm \delta y_{cp}$ & $\chi^2$ & DF\\
\hline
\hline
1. &\multicolumn{4}{|c|}{
$(x,y) \to (x,y)$}\\
\hline

    100. & $     10.14144 \pm       0.00155$ & $      0.47279 \pm       0.00021$ &    263.734 &    6 \\
    200. & $     13.95020 \pm       0.01307$ & $      0.47934 \pm       0.00047$ &     25.481 &    5 \\
    400. & $     19.40340 \pm       0.02031$ & $      0.48074 \pm       0.00064$ &     15.016 &    4 \\
    800. & $     27.03140 \pm       0.01527$ & $      0.48143 \pm       0.00076$ &     12.282 &    3 \\
   4000. & $     58.50040 \pm       0.08171$ & $      0.49023 \pm       0.00328$ &      4.686 &    2 \\
   8000. & $     82.54510 \pm       0.27698$ & $      0.47880 \pm       0.00702$ &      1.297 &    1 \\
  16000. & $    116.13100 \pm       1.05078$ & $      0.45799 \pm       0.01957$ &      0.000 &    0 \\
  32000. & $    159.52100 \pm       1.61276$ & & &\\

\hline
2. &\multicolumn{4}{|c|}{
$(x,y) \to (y,x) \vee (-y,-x)$}\\
\hline

    100. & $     10.14131 \pm       0.00197$ & $      0.47173 \pm       0.00015$ &   1916.220 &    6 \\
    200. & $     13.95890 \pm       0.00120$ & $      0.47619 \pm       0.00018$ &    221.267 &    5 \\
    400. & $     19.36230 \pm       0.04511$ & $      0.48387 \pm       0.00056$ &      7.323 &    4 \\
    800. & $     26.95380 \pm       0.00764$ & $      0.48408 \pm       0.00057$ &      3.333 &    3 \\
   4000. & $     58.61610 \pm       0.12235$ & $      0.48698 \pm       0.00192$ &      0.857 &    2 \\
   8000. & $     82.05480 \pm       0.24318$ & $      0.48904 \pm       0.00384$ &      0.470 &    1 \\
  16000. & $    114.94400 \pm       0.28547$ & $      0.49373 \pm       0.00783$ &      0.000 &    0 \\
  32000. & $    161.85000 \pm       0.78166$ & & &\\

\hline
3. &\multicolumn{4}{|c|}{
$(x,y) \to (-y,x) \vee (y,-x)$}\\
\hline
    100. & $     10.13721 \pm       0.00407$ & $      0.47885 \pm       0.00011$ &    584.104 &    6 \\
    200. & $     13.97810 \pm       0.00776$ & $      0.48137 \pm       0.00016$ &    149.121 &    5 \\
    400. & $     19.40490 \pm       0.02226$ & $      0.48388 \pm       0.00031$ &     57.209 &    4 \\
    800. & $     26.98040 \pm       0.02724$ & $      0.48550 \pm       0.00044$ &     30.702 &    3 \\
   4000. & $     58.54480 \pm       0.06015$ & $      0.49209 \pm       0.00129$ &      1.102 &    2 \\
   8000. & $     82.38670 \pm       0.02772$ & $      0.49019 \pm       0.00250$ &      0.322 &    1 \\
  16000. & $    115.68200 \pm       0.20951$ & $      0.49841 \pm       0.01471$ &      0.000 &    0 \\
  32000. & $    163.41900 \pm       1.63933$ & & &\\
\hline
\end{tabular}
\end{center}
\end{table}

\begin{table}[ht]
\protect\footnotesize
\begin{center}
\begin{tabular}{|c|c|c|c|c|}
\hline
4. & \multicolumn{4}{|c|}{
$(x,y) \to (-x,-y) $}\\
\hline
    100. & $     10.14555 \pm       0.00415$ & $      0.48117 \pm       0.00006$ &   2964.627 &    6 \\
    200. & $     13.94730 \pm       0.02649$ & $      0.48198 \pm       0.00006$ &     36.688 &    5 \\
    400. & $     19.34220 \pm       0.00095$ & $      0.48199 \pm       0.00006$ &     22.650 &    4 \\
    800. & $     27.00100 \pm       0.00421$ & $      0.48226 \pm       0.00010$ &     12.053 &    3 \\
   4000. & $     58.61070 \pm       0.12987$ & $      0.48721 \pm       0.00257$ &      8.353 &    2 \\
   8000. & $     81.96580 \pm       0.01518$ & $      0.49328 \pm       0.00432$ &      5.292 &    1 \\
  16000. & $    113.95800 \pm       0.70179$ & $      0.52218 \pm       0.01329$ &      0.000 &    0 \\
  32000. & $    163.65800 \pm       1.12058$ & & &\\

\hline
5. & \multicolumn{4}{|c|}{
$(x,y) \to (-x,y) \vee (x,-y)$}\\
\hline
    100. & $     10.14238 \pm       0.00359$ & $      0.47655 \pm       0.00021$ &    344.868 &    6 \\
    200. & $     13.93060 \pm       0.02251$ & $      0.48196 \pm       0.00037$ &     24.212 &    5 \\
    400. & $     19.39390 \pm       0.01713$ & $      0.48242 \pm       0.00041$ &     17.624 &    4 \\
    800. & $     26.97070 \pm       0.04447$ & $      0.48468 \pm       0.00073$ &      3.912 &    3 \\
   4000. & $     58.65150 \pm       0.07087$ & $      0.48689 \pm       0.00134$ &      0.034 &    2 \\
   8000. & $     82.21440 \pm       0.12537$ & $      0.48668 \pm       0.00231$ &      0.022 &    1 \\
  16000. & $    115.14900 \pm       0.34679$ & $      0.48752 \pm       0.00610$ &      0.000 &    0 \\
  32000. & $    161.44300 \pm       0.47969$ & & &\\

\hline
\hline
\end{tabular}
\end{center}
\caption{Average time spent in checking a failed cut--and---permute move
for different euivalence classes of moves. The check of the proposed configurations 
is performed as described in section~\ref{cc}}
\label{tabella_fail_cp}
\end{table}

\begin{table}[ht]
\begin{center}
\begin{tabular}{|c|c|c|c|c|}
\hline
\hline
$N$ & $\tau_{z^{end}_N} \pm \delta \tau_{z^{end}_N}$ & $z \pm \delta z$& $\chi^2$ &DF\\
\hline 
4000 &$10.347 \pm 0.040$& $ 0.1506 \pm  0.0029$&   1.564& 2 \\
8000 &$11.420 \pm 0.037$& $ 0.1543 \pm  0.0041$&  0.007&1 \\
16000 &$12.703 \pm 0.087 $& $0.155  \pm 0.012$&  0.000 & 0\\
32000 &$14.146 \pm   0.068 $&&&\\
\hline 
\end{tabular}
\end{center}
\caption{Autocorrelation times and dynamic exponent $z_{glob}$ for the observable $z^{end}$.}
\label{z}
\end{table}

\begin{thebibliography}{199}

\bibitem{napper}
D.~H.~Napper, {\em Polymeric Stabilization of Colloidal Dispersions}
(Academic, New York, 1983).

\bibitem{lip1}
R.~Lipowsky, Europhys. Lett. {\bf 30}, 197 (1995).

\bibitem{lip2}
R.~Hiergeist and R.~Lipowsky, J. Phys. II (France) {\bf 6}, 1465 (1996).

\bibitem{lip3}
R.~Lipowsky, H-G~D\"obereiner, C.~Hiergeist, and V.~Indrani,
Physica {\bf A 248(1-4)},536 (1998). 

\bibitem{franc}
V.~Frette, I.~Tsafir, M.-A.~Guedeau-Boudeville, L.~Jullien, D.~Kandel, and
J.~Stavans, Phys. Rev. Lett. {\bf 83}, 2465 (1999). 

\bibitem{Eisen_book}
E.~Eisenriegler, ``Polymers near surfaces", (World Scientific, Singapore) 1993.

\bibitem{Diehl_rev}
H.~W.~Diehl ``Field-theory of surface critical behaviour", in {\em Phase transitions
and and Critical Phenomena}, edited by C. Domb and J.L. Lebowitz, Vol. 10, p. 75.

\bibitem{Sokal1}
N.~Madras and  A.~D.~Sokal, 
J. Stat. Phys. {\bf 50}, 109 (1988).

\bibitem{Li}
B.~Li, N.~Madras, and A.~D.~Sokal,
J. Stat. Phys. {\bf 80}, 661 (1995).

\bibitem{Madras-Slade} 
N.~Madras and G.~Slade, {\it The Self-Avoiding Walk} (Birkh\"auser, 
Boston-Basel-Berlin, 1993)

\bibitem{Alan_Losanna}
A.~D.~Sokal, {\em Monte Carlo Methods for the Self-Avoiding Walk},
in Monte Carlo and Molecular Dynamics Simulations in Polymer Science,
K.~Binder editor, Oxford University Press (1994).

\bibitem{noi}
S.~Caracciolo, M.~S.~Causo, G.~Ferraro, M.~Papinutto, and A.~Pelissetto, 
J. Stat. Phys.{\bf 100}, 1111 (2000).

\bibitem{CPS-90}
S.~Caracciolo, A.~Pelissetto, and A.~D.~Sokal,
J. Stat. Phys. {\bf 60}, 1 (1990).

\bibitem{CPS-92}
S.~Caracciolo, A.~Pelissetto, and A.~D.~Sokal,
J. Stat. Phys. {\bf 67}, 65 (1992).

\bibitem{gamma}
S.~Caracciolo, M.~S.~Causo, and A.~Pelissetto, Phys. Rev. {\bf E 59}, R16 (1998).

\bibitem{grass}
R.~Hegger and P.~Grassberger, J. Phys. {\bf A 27}, 4069 (1994).

\bibitem{Diehl}
H.~W.~Diehl and M.~Shpot, Nucl. Phys. {\bf B 528}, 595 (1998).

\bibitem{diehl94} 
H.~W.~Diehl and M.~Shpot, Phys. Rev. Lett. {\bf 73}, 3431 (1994).

\bibitem{Kennedy}
T. Kennedy, J. Stat. Phys. {\bf 106}, 407 (2002).



\end{thebibliography}
\end{document}